\documentstyle[12pt]{article}

\renewcommand{\theequation}{\thesection.\arabic{equation}}

\newlength{\extraspace}
\newlength{\extraspaces}
\newcounter{dummy}

\newcommand{\baa}{
\addtocounter{equation}{1}
\setcounter{dummy}{\value{equation}}
\setcounter{equation}{0}
\renewcommand{\theequation}{\thesection.\arabic{dummy}\alph{equation}}
\begin{eqnarray}
\addtolength{\abovedisplayskip}{\extraspaces}
\addtolength{\belowdisplayskip}{\extraspaces}
\addtolength{\abovedisplayshortskip}{\extraspace}
\addtolength{\belowdisplayshortskip}{\extraspace}}

\newcommand{\eaa}{
\end{eqnarray}
\setcounter{equation}{\value{dummy}}
\renewcommand{\theequation}{\thesection.\arabic{equation}}}

\newcommand{\be}{\begin{equation}
\addtolength{\abovedisplayskip}{\extraspaces}
\addtolength{\belowdisplayskip}{\extraspaces}
\addtolength{\abovedisplayshortskip}{\extraspace}
\addtolength{\belowdisplayshortskip}{\extraspace}}
\newcommand{\ee}{\end{equation}}

\newcommand{\ba}{\begin{eqnarray}
\addtolength{\abovedisplayskip}{\extraspaces}
\addtolength{\belowdisplayskip}{\extraspaces}
\addtolength{\abovedisplayshortskip}{\extraspace}
\addtolength{\belowdisplayshortskip}{\extraspace}}
\newcommand{\ea}{\end{eqnarray}}

\newcommand{\bd}{\begin{displaymath}
\addtolength{\abovedisplayskip}{\extraspaces}
\addtolength{\belowdisplayskip}{\extraspaces}
\addtolength{\abovedisplayshortskip}{\extraspace}
\addtolength{\belowdisplayshortskip}{\extraspace}}
\newcommand{\ed}{\end{displaymath}}

\newcommand{\ban}{\begin{eqnarray*}
\addtolength{\abovedisplayskip}{\extraspaces}
\addtolength{\belowdisplayskip}{\extraspaces}
\addtolength{\abovedisplayshortskip}{\extraspace}
\addtolength{\belowdisplayshortskip}{\extraspace}}
\newcommand{\ean}{\end{eqnarray*}}

\newcommand{\newsection}[1]{
\vspace{15mm}
\pagebreak[3]
\addtocounter{section}{1}
\setcounter{equation}{0}
\setcounter{subsection}{0}
\setcounter{footnote}{0}
\begin{center}
{\Large \thesection. #1}
\end{center}
\nopagebreak
\medskip
\nopagebreak}

\newcommand{\newsubsection}[1]{
\vspace{1cm}
\pagebreak[3]

\addtocounter{subsection}{1}
\noindent{ \sc \thesubsection. #1}
\nopagebreak
\vspace{2mm}
\nopagebreak}

\newcommand{\nonu}{\nonumber \\[.5mm]}

\newcommand{\deel}[2]{{\textstyle{#1 \over #2}}}
\newcommand{\hf}{{\textstyle{1\over 2}}}
\newcommand{\hv}{{\textstyle{1\over 4}}}

\newcommand{\ie}{{\it i.e.}}

\newtheorem{lemma}{Lemma}
\newtheorem{theorem}{Theorem}
\newtheorem{thm}{Theorem}
\newtheorem{lmm}{Lemma}
\newtheorem{exam}{Example}
\newcommand{\bt}{\begin{thm}}
\newcommand{\et}{\end{thm}}
\newcommand{\bl}{\begin{lmm}}
\newcommand{\el}{\end{lmm}}
\newcommand{\bex}{\begin{exam}}
\newcommand{\eex}{\end{exam}}
\newcommand{\hj}{\hat{J}}
\newcommand{\whj}[1]{W(\hat{J}^{#1})}
\newcommand{\www}[4]{\deel{#1}{#2}\hj^{#3}\hj^{#4}}

\def\inbar{\,\vrule height1.5ex width.4pt depth0pt}
\font\rms=cmr12 at 12pt
\def\ce{\relax\ifmmode\mathchoice
{\hbox{$\inbar\kern-.3em{\rm C}$}}
{\hbox{$\inbar\kern-.3em{\rm C}$}}
{\lower.9pt\hbox{\rms $\inbar\kern-.3em{\rm C}$}}
{\lower1.2pt\hbox{\rms $\inbar\kern-.3em{\rm C}$}}
\else{$\inbar\kern-.3em{\rm C}$}\fi}
\font\cmss=cmss12 \font\cmsss=cmss12 at 12pt
\def\ze{\relax\ifmmode\mathchoice
{\hbox{\cmss Z\kern-.4em Z}}{\hbox{\cmss Z\kern-.4em Z}}
{\lower.9pt\hbox{\cmsss Z\kern-.4em Z}}
{\lower1.2pt\hbox{\cmsss Z\kern-.4em Z}}\else{\cmss Z\kern-.4em Z}\fi}

\newcommand{\tr}{\mbox{Tr}}

\newcommand{\actie}[1]{\deel{1}{2\pi}\int d^2z \, }

\newcommand{\mats}[9]{\left( \begin{array}{ccc}
				#1 & #2 & #3 \\
				#4 & #5 & #6 \\
				#7 & #8 & #9
                             \end{array} \right) }

\newcommand{\np}[1]{Nucl. Phys. {\bf B#1}}
\newcommand{\cmp}[1]{Comm. Math. Phys. {\bf #1}}

\newcommand{\plb}[1]{Phys. Lett. {\bf B#1}}

\newcommand{\ad}[1]{\mbox{\rm ad}_{#1}}

\newcommand{\im}{\mbox{\rm im}}

\begin{document}
\addtolength{\baselineskip}{.7mm}

\thispagestyle{empty}
\begin{flushright}
{\sc THU}-92/32\\
{\sc ITFA}-28-92
\end{flushright}
\vspace{1.5cm}
\setcounter{footnote}{2}
\begin{center}
{\LARGE\sc{Quantization and
representation theory of finite W algebras}}\\[0.8cm]

\sc{Jan de Boer\footnote{e-mail: deboer@ruunts.fys.ruu.nl}}
\\[1.5mm]
{\it Institute for Theoretical Physics\\[0.7mm]
Princetonplein 5\\[0.7mm]
P.O. Box 80.006\\[0.7mm]
3508 TA Utrecht\\[6.5mm]}
\sc{ Tjark Tjin \footnote{e-mail: tjin@phys.uva.nl}}
\\[1.5mm]
{\it Institute for Theoretical Physics\\[0.7mm]
Valckenierstraat 65\\[0.7mm]
1018 XE Amsterdam\\[0.7mm]
The Netherlands\\[6mm]}

\message{baselineskip=\the\baselineskip}

{\sc Abstract}\\[0.5cm]
\end{center}

\noindent
{\footnotesize \baselineskip 10pt
In this paper we study the finitely generated algebras underlying
$W$ algebras. These so called 'finite $W$ algebras' are constructed as
Poisson reductions of  Kirillov Poisson structures on
simple Lie algebras. The inequivalent reductions
are labeled by the inequivalent embeddings of $sl_2$ into the
simple Lie algebra in question. For arbitrary embeddings a
coordinate free formula for the reduced Poisson structure is derived.
We also prove that any finite $W$ algebra can be embedded
into the Kirillov Poisson algebra of a (semi)simple Lie algebra
(generalized Miura map). Furthermore it is shown that generalized finite
Toda systems are reductions
of a system describing a free particle moving
on a group manifold and that they have finite $W$ symmetry.
In the second part we BRST quantize the finite
$W$ algebras.
The BRST cohomology is calculated using a spectral
sequence (which is different from the one used
by Feigin and Frenkel). This allows us to quantize all finite
$W$ algebras in one stroke. Explicit results for $sl_3$ and $sl_4$
are given. In the last part of the paper we study the representation
theory of finite $W$ algebras. It is shown, using a quantum version
of the generalized Miura
transformation, that the representations of finite $W$ algebras
can be constructed from the representations of a certain Lie subalgebra
of the original simple Lie algebra. As a byproduct of this we are able
to construct the Fock realizations of arbitrary finite $W$ algebras.

\message{baselineskip=\the\baselineskip}}

\vfill

\newpage

\newsection{Introduction}

It is only relatively recent that it was realized that nonlinear
symmetry algebras play an important role in physics. The discovery
of $W$ algebras in Conformal Field theory \cite{Za}
(see \cite{BoSc} for a recent review) made it clear
that they would play an important role in string theory, field theory,
integrable systems and the theory of 2D critical phenomena. One reason
for their late discovery is that up to now they are only known as
infinitesimal symmetries. The global invariances of a system with a
nonlinear symmetry are not known. This is of course related to the fact
that nonlinear algebras don't exponentiate to groups like Lie algebras.

A lot of work has been done on trying to understand what the meaning
of $W$ algebras is. It turns out that many $W$ algebras found in CFT
are not completely unrelated to the linear theory of Lie algebras
after all. This was first realized when it was shown in \cite{BaMa}
that certain Poisson algebras occurring in the theory of integrable
hierarchies of evolution equations were nothing but classical versions
of the nonlinear algebras found in CFT. In particular the well known
$W_n$ algebras are related in this way to the second Hamiltonian structures
of KdV like hierarchies. These hierarchies, and their Hamiltonian
structures were however already shown to be reductions of a different
class of integrable hierarchies which have a second Hamiltonian structure
that is equal to the Kirillov Poisson structure on the dual of an affine
Lie algebra \cite{DS}. This means on an algebraic level that classical
$W_n$ algebras can be obtained from affine Lie algebras by Poisson
reduction. This picture was worked out in \cite{BFFOW} where it was
shown that a classical $W_n$ algebra is nothing but the Dirac bracket
algebra on a submanifold of the affine Lie algebra.

In the meantime many new $W$ algebras were constructed by what can be
called the 'direct method', i.e. by imposing Jacobi identities on
general nonlinear extensions of the Virasoro algebra. Since the Jacobi
identities are themselves nonlinear algebraic equations the construction
of $W$ algebras in this manner is rather cumbersome. It was therefore
a natural question to ask (also from the point of view of Poisson
reduction of Poisson manifolds) whether more of these algebras
can be obtained via Poisson reduction from affine Lie algebras.
That this is the case was shown in \cite{BTV} where the construction
of \cite{DS,BFFOW} was generalized to include many more $W$ algebras
besides $W_n$. Motivated by a similar situation encountered in the
theory of dimensional reductions of selfdual Yang-Mills equations
it was shown that to every embedding of $sl_2$ into the simple Lie
algebra underlying the affine algebra there is associated a Poisson
reduction leading to a $W$ algebra. $sl_2$ embeddings that are related
to one another by inner automorphisms lead to the same reductions, so
in order to find out how many inequivalent reductions there are one
needs to count the number of equivalence classes of $sl_2$ embeddings.
For $sl_n$ this number is $P(n)$, the number of partitions of $n$.
The standard reductions leading $W_n$ algebras turned out
to be associated
to the so called 'principal embeddings'.

The fact that one knows that these $W$ algebras have a linear origin
helps a lot when one tries to analyse them. For example the construction
of invariant chiral actions is facilitated as was shown in \cite{Dublin}.
Also the construction of the classical covariant $W$ gravities and
their moduli spaces have been made possible by this \cite{BG,Wgrav}.

The procedure of Poisson reduction is of course purely classical. In
order to really make contact  with CFT one would like to quantize
the $W$ algebras obtained by Poisson reduction. The $W_n$ algebras
were quantized in \cite{FL} by (naively) quantizing the well known
Miura transformation. In essence what one does is classically express
the $W_n$ generators in terms of classical harmonic oscillators
via the Miura transform. One then quantizes the $W_n$ algebra by
quantizing the harmonic oscillators and normal ordering. More or less
by accident this gives a quantum algebra that closes for $A_n$. It does
not work for all algebras however \cite{BoSc},
as was to be expected since this is
in general not a valid quantization procedure. The quantization
of the Poisson reduction leading to $W_n$ algebras was made more
precise in \cite{FF} where the BRST formalism was used to tackle this
problem. In order to calculate the BRST cohomology Feigin and Frenkel
proposed to use spectral sequences. The paper \cite{FF} is however
incomplete as we shall see in this paper and furthermore they only
consider the reductions associated to the principal embeddings.

It was also attempted to construct the representation theory of $W$
algebras form the representation theory of affine Lie algebras
\cite{FKW,FF}. In principle the BRST procedure provides a
functor from the representation theory of affine Lie algebras
to the representation theory of $W$ algebras. It turned out
to be possible to obtain $W_n$ characters from Kac-Moody characters.
The general theory of $W_n$ representations is however far from
complete. Furthermore for the other reductions the quantizations
(let alone the representation theories) are not known.

Up to now the study of $W$ algebras has concentrated on the infinite
dimensional case. This situation is comparable to trying to develop
the theory of Lie algebras by starting with the infinite dimensional
case. As the structure and representation theory of affine Lie algebras
is largely determined by that of the finite dimensional simple Lie
algebras that underly them it is our opinion that it might be helpful
to look for and study the finitely generated structures underlying
$W$ algebras. This program was initiated in \cite{tjark} and will
be carried out in the present paper. It will turn out that the theory
of 'finite $W$ algebras' is remarkably rich and contains already many
of the features encountered in ordinary $W$ algebras. It is therefore
our expectation that much of what we will say in this paper will
transfer without much alteration to the infinite dimensional case.

The paper is roughly split up into three parts. The first part
deals with the classical theory. Classical finite $W$ algebras are
constructed as Poisson reductions of Kirillov Poisson structures
on simple Lie algebras (in complete analogy with ordinary $W$ algebras
which are constructed as reductions of affine Lie algebras). The
Poisson algebras thus obtained are nonlinear and finitely generated.
We discuss their structure and show that in general they do have
linear Poisson subalgebras that are isomorphic to Kirillov Poisson
algebras. We also derive a coordinate free expression for the reduced
Poisson structure of an arbitrary reduction. The Miura transformation
turns out to have a finite dimensional analogue  which can in fact
be extended to arbitrary reductions. From this it follows that any
finite $W$ algebra can be embedded into the Kirillov Poisson algebra
of a certain subalgebra of the simple Lie algebra with which we started.
At the end of the first part of the paper we investigate  which theories
have finite $W$ symmetry. It turns out that  (as could have been
expected) these are generalized finite Toda systems. In deriving
this however we show that finite Toda systems are reductions
of a system describing a free particle moving on a group manifold.
This allows us to give general formulas for the solution space
of such systems.

In the second part of the paper we BRST quantize the finite $W$ algebras.
The nontrivial part of this is of course calculating the BRST cohomology
and its algebraic structure. Since the BRST differential is a sum of
two other differentials
one can associate a double complex to the BRST complex.
In order to calculate the BRST cohomology one can then use the theory
of spectral sequences.
There is a choice to be made between one out of
two spectral sequences that one can associate to a double complex. These
spectral sequences must give the same final
answer for the BRST cohomology,
as is well known from the theory of spectral sequences,
but for the calculation it is
crucial which one one takes. The choice we make is different from
the one made by Feigin and Frenkel and allows us to quantize any
finite $W$ algebra and reconstruct its algebraic structure. In order
to  make our construction  more explicit we calculate all finite
quantum $W$ algebras that can be obtained from $sl_2$, $sl_3$ and $sl_4$.

In the third and last part of the paper we discuss the representation
theory of finite $W$ algebras. Crucial for this is a quantum version
of the generalized Miura transformation which embeds any finite
$W$ algebra into the universal enveloping algebra of some (semi)simple
Lie algebra. An arbitrary representation of this Lie algebra
therefore immediately yields a representation of the finite $W$
algebra. This also allows us to derive Fock realizations for arbitrary
finite $W$ algebras since Fock realizations for simple Lie algebras
are well known. This replaces the cumbersome construction of $W$
algebras as commutants of screening operators. As an illustrative
example we realize the finite dimensional representations of the
finite $W$ algebra $\bar{W}_3^{(2)}$ as a subrepresentation of certain Fock
realizations. In principle this provides the first term of a Fock space
resolution of these representations \cite{BMP1}.

We have tried to keep the paper as self-contained as possible and
give full proofs of the main assertions.

\newsection{Basics of Kirillov Poisson structures and Poisson reduction}

In order to make the paper reasonably selfcontained we briefly discuss
in this section the Kirillov Poisson structure on a Lie algebra and
Poisson reduction of Poisson manifolds. These two concepts will then
be put together in the remainder of the paper.

\newsubsection{Kirillov Poisson structures}

Let $(M,\{.,.\})$ be a Poisson manifold,
that is $\{.,.\}$ is a Poisson bracket on the space
$C^{\infty}(M)$ of $C^{\infty}$ functions on $M$,
and $G$ a Lie group. Also let
$\Phi :G\times M \rightarrow M$ be a smooth and proper
action of $G$ on $M$ which
preserves the Poisson structure, i.e.
\begin{equation}
\Phi_g^*\{\phi , \psi \} = \{\Phi_g^*(\phi ), \Phi_g^*(\psi )\}
\end{equation}
where $\Phi^*_g$ is the pullback of $\Phi_g:M \rightarrow M$.
Physically this implies that if $\gamma (t)$ is a solution of the
equations of motion w.r.t. some $G$-invariant Hamiltonian H (i.e.
$\Phi_g^* H=H$  where as usual $\Phi^*_gH=H \circ \Phi_g$), then
$(\Phi_g \circ \gamma )(t)$ is again a solution. If we do not want
to consider solutions that can be transformed into each other in this
way to be essentially different we are led to consider the space
$M/G$ as the 'true' phase space of the system. The only observables
(i.e. functions on $M$) that one is concerned with are those which are
themselves G-invariant and therefore descend to observables on
$M/G$. Denote the set of smooth $G$-invariant functions on $M$ by $O$,
and let then $\phi, \psi \in O$. Using the fact that $\Phi_g$ preserves
the Poisson structure for all $g \in G$ and that $\phi$ and $\psi$
are $G$-invariant we can easily deduce that $\{\phi,\psi\}$ is also
an element of $O$, i.e. $O$ is a Poisson subalgebra of $C^{\infty}(M)$.
Let $\pi : M \rightarrow M/G$ be the canonical projection. The
pullback map $\pi^* : C^{\infty}(M/G) \rightarrow O$ is in fact
an isomorphism. It assigns to a function $\hat{\phi}$ on $M/G$
the function $\hat{\phi}\circ \pi$ on $M$ which is constant along
the $G$-orbits. What one wants is now to define a Poisson structure
$\{.,.\}^*$ on $C^{\infty}(M/G)$ such that the Poisson algebras
$(C^{\infty}(M/G), \{.,.\}^*)$ and $(O,\{.,.\})$ are isomorphic.
This would mean that all ($G$-invariant) information of the original
phase space is transferred to the Poisson algebra $(C^{\infty}(M/G),
\{.,.\}^*)$. Therefore define
\begin{equation}
\{ \hat{\phi},\hat{\psi}\}^*=(\pi^*)^{-1}\{\pi^*\hat{\phi},
\pi^*\hat{\psi}\}
\end{equation}
for all $\hat{\phi},\hat{\psi} \in C^{\infty}(M/G)$.
Obviously $\pi^*$ is a Poisson algebra isomorphism between
the two Poisson algebras. So starting from a $G$-invariant theory
we have arrived at a formulation in which all essential information
is contained and all redundancy has been eliminated.

Consider now a $G$-invariant theory in which the group manifold itself
is the configuration space (for example a particle moving on $G$ with
a Hamiltonian that is invariant under $G$). The phase space of such
a system is of course $T^*G$, the cotangent bundle of $G$. The left action
of the group on itself, denoted by $L_g$,  induces a left $G-$action
$L_g^*$ on $T^*G$ which is in fact a Poisson action w.r.t. the
canonical Poisson structure on $T^*G$ ($T^*G$ carries a natural Poisson
structure because it is a cotangent bundle). Therefore we can do what
we did before and extract the $G$-invariant piece of the theory by
going to the quotient manifold
\begin{equation}
T^*G/G
\end{equation}
Obviously this space is nothing but the space of left invariant
one forms which means that it is isomorphic to $g^*$, the dual of
the Lie algebra of $G$. The canonical projection
\begin{equation}
\pi:T^*G \rightarrow g^*
\end{equation}
is therefore simply the pullback to the fiber over the identity element
of $G$ by means of left translation. As before the Poisson structure on
$T^*G$ induces a Poisson structure  on $g^*$. This Poisson structure
on $g^*$ is called the 'Kirillov' Poisson structure.

It is possible to obtain (by explicit calculation) a more concrete
formula for the Kirillov Poisson structure \cite{AbMa}.
It reads
\begin{equation}
\{F,G\}(\xi )=\xi ([\mbox{grad}_{\xi}F,\mbox{grad}_{\xi}G])
\end{equation}
where $F$ and $G$ are smooth functions on $g^*$, $\xi \in g^*$ and
$\mbox{grad}_{\xi}F \in g$ is determined by
\begin{equation}
\frac{d}{d\epsilon}F(\xi+ \epsilon \eta)|_{\epsilon =0}
=\eta(\mbox{grad}_{\xi}F)
\end{equation}
for all $\eta \in g^*$. In  this paper we will always take $G$ to be
a simple group which means that the Cartan-Killing form $(.,.)$
on its Lie algebra is non-degenerate. This allows us to identify
the Lie algebra $g$ with its dual, i.e. all $\xi \in g^*$ are of
the form $(\alpha,.)$ where $\alpha \in g$. The above formulas then
become
\begin{equation}
\{F,G\}(\alpha )=(\alpha,[\mbox{grad}_
{\alpha}F,\mbox{grad}_{\alpha}G])  \label{kir}
\end{equation}
and
\begin{equation} \label{df}
\frac{d}{d\epsilon}F(\alpha + \epsilon \beta) |_{\epsilon =0}=
(\beta,\mbox{grad}_{\alpha}F)
\end{equation}
where $\alpha,\beta \in g$ and $F,G \in C^{\infty}(g)$. This
is the form that we will use.

It is possible to give an even more detailed description of the
bracket by choosing a basis $\{t_a\}$ in $g$. Define the function
$J^a$ on $g$ by $J^a(t_b)=\delta^a_b$ and extend it linearly to all
of $g$. Since $J^a(\alpha + \epsilon\beta)=\alpha^a+\epsilon \beta^a$,
where $\alpha^a$ are the components of $\alpha$ in the basis $\{t_a\}$,
we find that the left hand side of eq.(\ref{df}) is equal to $\beta^a$.
Now, one can write $\mbox{grad}_{\alpha}J^a=
(\mbox{grad}_{\alpha}J^a)^bt_b$ which means that we find for the
right hand side $\beta^c(\mbox{grad}_
{\alpha}J^a)^bg_{cb}$ where $g_{cb}=
(t_c,t_b)$. Eq.(\ref{df}) therefore reads
\begin{equation}
\beta^c(\mbox{grad}_{\alpha}J^a)^bg_{cb}=\beta^a
\end{equation}
which means that
\begin{equation}
(\mbox{grad}_{\alpha}J^a)^b=g^{ab}
\end{equation}
where $g^{ab}$ is the inverse of $g_{ab}$. Inserting this into
eq.(\ref{kir}) we find
\begin{equation}
\{J^a,J^b\}=f^{ab}_{c}J^c    \label{kir2}
\end{equation}
where $f_{ab}^c$ are the structure constants of $g$ in the basis
$\{t_a\}$. The resemblance with the Lie bracket
\begin{equation}
[t_a,t_b]=f_{ab}^ct_c
\end{equation}
on $g$ is striking however we have to remember that the Poisson bracket
(\ref{kir2}) is defined on the space of smooth functions on $g$ which
is a Poisson algebra, i.e. it also carries a commutative multiplication
compatible with the Poisson structure . This means for example that
the expression $J^aJ^bJ^c$ makes sense, while $t_at_bt_c$ does not
(remember that if $F,G$ are two functions on $g$ then their product
is defined by $(FG)(\alpha)=F(\alpha)G(\alpha)$).

In this paper we shall be interested in certain Poisson reductions
of the Poisson structure discussed above. Let us therefore briefly
review the procedure called Poisson reduction.

\newsubsection{Poisson reduction}

In this section we briefly describe the procedure called 'Poisson
reduction'. Our presentation does not pretend to be complete or rigorous,
but is a heuristic introduction to the subject. For more
detailed discussions  we refer to \cite{Poiss}.

Let $(M,\{.,.\})$ be a Poisson manifold and
let $\{\phi_i\}_{i=1}^q$ be a set of independent
elements of $C^{\infty}(M)$. Denote by $C$ the submanifold
\be
C=\{p \in M \mid \phi_i (p)=0 \mbox{ for all }i=1, \ldots ,q\}
\ee
and let $j:C \hookrightarrow M$ be the canonical embedding.
As usual we call the functions $\{\phi_i\}$ constraints and say that
$\phi_i$ is 'first class' if on $C$ (i.e. after imposing the constraints)
it Poisson commutes with all other constraints. A constraint is called
second class if it is not first class.

It is a natural question to ask whether the Poisson bracket $\{.,.\}$
on $M$ induces a Poisson bracket on $C$. In general the answer to this
question is no. In order to see this let us assume that the Poisson
bracket $\{.,.\}
$ is derived from a symplectic form $\omega$ on $M$ (in general
$M$ is the union of symplectic leaves).The symplectic form
$\omega$ then induces a 2-form $j^*\omega$ on $C$ which is nothing
but the pull back. This 2-form is closed, for we have
\be
d(j^*\omega )=j^*d\omega  =0
\ee
but the trouble is that it may no longer be non-degenerate. This
phenomenon is caused by the first class constraints. In order to see
this, let $X_i$ be the Hamiltonian vectorfield associated to the
first class constraint $\phi_i$, i.e.
\be
i_{X_i}\omega=d \phi_i
\ee
Since $X_i|_{C}(\phi_j)\equiv \{\phi_i,\phi_j\}(C)=0$ for all $j$ we find
that $X_i|_C$ is tangent to $C$. Now, let $Y$ be any vectorfield
on $M$ tangent to $C$, then
\begin{eqnarray}
(j^*\omega )(X_i|_C,Y|_C) & \equiv & \omega (X_i,Y)|_C \nonumber \\
& = & Y(\phi_i )|_C \nonumber \\
& = & 0
\end{eqnarray}
where in the first step we used that both $X_i$ and $Y$ are tangent
to $C$ and in the last step that $\phi_i$ is constant  on $C$. We
see that the Hamiltonian vectorfields of the first class constraints
are in the kernel of the 2-form $j^*\omega$ which is therefore
degenerate. In physical terms what happens
here is that the first class constraints generate gauge invariances
on $C$, i.e. a curve $\gamma (t)$ on $C$ tangent to the vectorfield
$X_i$ (which was associated to a first class constraint) consists of
physically equivalent 'states'. What we need to do is identify all these
states. In the systems we will consider the Hamiltonian vectorfields
of the first class constraints always form an involutive system which
means, by Frobenius' theorem, that $C$ splits up into a union of
nonintersecting submanifolds (leaves). These leaves are
sets of physically equivalent states and their tangent spaces coincide
with the degeneracy of the form $j^*\omega$. It is clear that in order
to get a symplectic form one needs to identify points that lie on the
same leaf. Denote the resulting coset space by $\bar{M}$ and let
$\pi :C \rightarrow \bar{M}$ be the canonical projection. The symplectic
form $\bar{\omega}$ on $\bar{M}$ is then defined by \cite{AbMa}
\be
j^*\omega=\pi^*\bar{\omega}
\ee
For the systems considered in this paper the flows of the Hamiltonian
vectorfields of the first class constraints are complete and
generate a (gauge) group $H$. The triple $(C, \bar{M},H)$ is then
a principal bundle with total space $C$, base space $\bar{M}$ and
structure group $H$. Assuming that this bundle is trivial
(as will always be the case) we can choose a global section $s$
(in physics terminology this corresponds to choosing a gauge). The
composite map $j \circ s$ is then an embedding of $\bar{M}$ into $M$.
There is now a simple relation between $\bar{\omega}$  and $\omega$,
namely
\be
\bar{\omega}=(\pi \circ s)^*\bar{\omega} =s^*\pi^* \bar{\omega}
=s^*j^*\omega
=(j\circ s)^* \omega
\ee
since $\pi \circ s =1$ by definition of a section.

Let us choose a section $s$ of $C$ such that there exist 'gauge fixing
constraints' $\phi_{q+1}, \ldots, \phi_n$ with the property that
\be
j \circ s (\bar{M})=\{p \in M \mid \phi_i(p)=0 \mbox{ for } i=1, \ldots,
n\}
\ee
We can give the relation between the Poisson bracket on $M$ induced
by $\omega$ and the Poisson bracket $\{.,.\}^*$ on $\bar{M}$ induced
by $\bar{\omega}$ in an explicit way. It reads
\be
\{\bar{f},\bar{g}\}^*=\overline{\{f,g\}-\sum_{ij=1}^{n}\{f,\phi_i\}
\Delta^{ij}\{\phi_j,g\}} \label{di}
\ee
where $f,g \in C^{\infty}(M)$, the bar denotes the restriction
to $\bar{M}$ and $\Delta^{ij}$ is the inverse of
$\Delta_{ij}=(\{\phi_i,\phi_j\})_{i,j=1}^n$. The
proof of this formula will not be given here but can be found in
\cite{Sund}. The bracket (\ref{di}) was originally discovered by Dirac
and is therefore called the 'Dirac bracket'. One can immediately see
from (\ref{di}) that the Dirac bracket is not defined if any of the
constraints $\phi_i$ are first class, for then $\bar{\Delta}_{ij}$
would have at least one column with only zeros which means that
$\Delta$ would no longer be invertible. This corresponds to the earlier
statement that in the presence of first class constraints $j^*\omega$
is not
symplectic.

Everything we have said above was in the terminology of symplectic
geometry, but from (\ref{di}) follows that the Dirac bracket is still
defined if $\{.,.\}$ is a Poisson structure that is not symplectic.
The crucial thing is that after constraining and gauge fixing we
are left with a set of second class constraints.

\newsection{Classical Finite $W$ algebras}

In this section we apply the theory outlined  in the previous
section to the Kirillov Poisson structure. It will be shown that
every $sl_2$ embedding determines in a natural way a Poisson
reduction of the Kirillov Poisson structure. The resulting Poisson
algebras are called 'classical Finite $W$ algebras'.
We start with a brief review of the theory of $sl_2$ embeddings
into a simple Lie algebra \cite{Dynkin}.

\newsubsection{$sl_2$-embeddings}

Let $i: sl_2 \hookrightarrow g$ be an embedding of $sl_2$ into a
simple Lie algebra $g$
(in what follows $g$ will always be
$sl_n$), and let $L(\Lambda ;n )$ be the irreducible finite
dimensional representation of $g$
with highest weight $\Lambda$. As a representation of the
subalgebra $i(sl_2)$  of $g$,  $L(\Lambda;n )$ need not
be irreducible. In general it decomposes into a
direct sum of irreducible $sl_2$ representations
\begin{equation}
L(\Lambda ;n) \simeq \bigoplus_{j \in \frac{1}{2} \bf N} n_j(
\Lambda;i) \cdot
\underline{2j+1}_2
\end{equation}
where $\underline{2j+1}_2$ denotes the $(2j+1)$-dimensional irreducible
representation of $sl_2$ and $n_j(\Lambda ;i)$
are the multiplicities of these
representations in the decomposition.
The set of numbers $\{n_j(\Lambda;i)\}_j$ is called the
'branching rule' of the representation $L(\Lambda ;n)$
under the $sl_2$ embedding
$i$. In what follows $L(\Lambda ;n)$ will always either be the
($n-$dimensional) fundamental
or the ($n^2-1$ dimensional)
adjoint representation of $sl_n$ which we simply denote by
$\underline{n}_n$ and
$\underline{ad}_n$ respectively.

Two $sl_2$ embeddings $i$ and $i'$ are called equivalent if there
exists an inner automorphism $\psi $ of $g$ such that
\begin{equation}
i'=\psi \circ i
\end{equation}
It is easy to show that if $i \sim i'$ then $n_j(\Lambda;i)=
n_j(\Lambda;i')$
for all representations $L(\Lambda ;n)$ and $j
\in \frac{1}{2} \bf N$. The converse is
not so clear. However, for $g=sl_n$ it can
be shown  that \cite{Dynkin}
\begin{enumerate}
\item The inequivalent $sl_2$ embeddings into $sl_n$ are completely
characterized by the branching rule of the fundamental representation,
i.e. two $sl_2$ embeddings are equivalent if and only if the
branching rules of the fundamental
representation w.r.t. these two embeddings are equal.
\item There exist $P(n)$ inequivalent $sl_2$ embeddings into $sl_n$,
where $P(n)$ is the number of partitions of the number $n$.
\end{enumerate}
The set of equivalence classes of $sl_2$ embeddings
into $sl_n$ is therefore
parametrized by the branching rules of the fundamental representation,
while the different branching rules that are possible are simply given
by the partitions of $n$.\\[2mm]

{\it Example:
Let $g=sl_4$. There are 5 equivalence classes of $sl_2$ embeddings
into $sl_4$ characterized by the following decompositions
\begin{enumerate}
\item $\underline{4}_4 \simeq \underline{4}_2$
\item $\underline{4}_4 \simeq \underline{3}_2 \oplus \underline{1}_2$
\item $\underline{4}_4 \simeq 2 \cdot \underline{2}_2$
\item $\underline{4}_4 \simeq \underline{2}_2 \oplus 2 \cdot \underline{1}_2$
\item $\underline{4}_4 \simeq 4 \cdot \underline{1}_2$
\end{enumerate}
of the fundamental representation $\underline{4}_4$ of $sl_4$.
}\\[2mm]

The embeddings corresponding to the first branching rule of this
example are called 'principal'. In general they are characterized
by the fact that the fundamental representation becomes an irreducible
representation of the embedded algebra. The last branching rule of the
example corresponds to the so called 'trivial embeddings'. They are
characterized by the fact that only singlets appear in the decomposition
of the fundamental representation.

In what follows  we shall also need the centralizer $C(i)$ of
$i(sl_2)$ in $g$. It corresponds to the singlets of
the adjoint action of $i(sl_2)$ in $g$.
In order to
describe $C(i)$ we have to take a look at the branching rule of the
fundamental representation
\begin{equation}
\underline{n}_n \simeq \bigoplus_{j \in \frac{1}{2}
\bf N} n_j
\cdot \underline{2j+1}_2
\end{equation}
where for convenience we simply denote $n_j \equiv n_j(\Lambda;i)$
for the fundamental representation.
Let
$q$ be the number of different values for $j$ appearing in this
decomposition for which $n_j$ is non-zero, then $C(i)$ is given by
\begin{equation}
C(i)\simeq \bigoplus_{j \in \frac{1}{2} \bf N}sl_{n_j} \; \bigoplus
(q-1)u(1)
\end{equation}
so in essence it is a direct sum of $sl_{n_j}$ algebras.

{}From the branching of the fundamental representation
one can deduce
the branching of the $(n^2-1)$-dimensional adjoint representation
$\underline{ad}_n$ of $g=sl_n$. It reads
\begin{eqnarray}
\underline{ad}_n \oplus \underline{1}_n
& \simeq & \bigoplus_{j\in \bf N} n_j^2 \cdot
(\underline{1}_2
\oplus \underline{3}_2 \oplus \ldots \oplus
\underline{4j+1}_2) \nonumber \\
& & \bigoplus_{j \neq j'} n_jn_{j'}\cdot
(\underline{2|j-j'|+1}_2 \oplus \ldots
\oplus \underline{2|j+j'|+1}_2)
\end{eqnarray}
where the singlet on the left hand side was added  to indicate
that in order to obtain the branching rule of the adjoint representation
$\underline{ad}_n$ we still have to 'subtract' a singlet from the
righthand side.

Let $\{t_0,t_+,t_-\}$ be the standard
generators of $i(sl_2)$. The Cartan element $t_0$, called the
defining vector of the embedding, can always be chosen to be
an element of the Cartan subalgebra of $g$ \cite{Dynkin}. Therefore
it defines a $\frac{1}{2}\bf Z$  gradation of $g$
given by
\be
g=\bigoplus_{m \in \frac{1}{2} \bf Z}g^{(m)}\;\; ; \;\;\;\;\
\ee
We can choose a basis
\be
\{t_{j,m}^{(\mu )}\}_{j \in \frac{1}{2}{\bf N}; \;
-j \leq m \leq j; \; 1 \leq \mu \leq n_j}
\ee
for $g=sl_n$ such that
\begin{eqnarray}
[t_3,t_{j,m}^{(\mu )}] & = & m t_{j,m}^{(\mu )} \nonumber \\ {}
[t_{\pm},t_{j,m}^{(\mu )}] & = & c(j,m)t_{j,m\pm 1}^{(\mu )}
\end{eqnarray}
where $c(j,m)$ are standard normalization factors.
We will always take the labeling to be such that $t^{(1)}_{1,\pm 1}\equiv
t_{\pm}$ and $t_{1,0}^{(1)} \equiv t_0$.
It is clear that
\begin{equation}
g^{(m)}= \bigoplus_{j,\mu} {\bf C} t_{j,m}^{(\mu)}
\end{equation}
Note that the spaces $g^{(m)}$ and $g^{(-m)}$ are always of the same
dimension. We have the following
\begin{lemma}
The spaces $g^{(m)}$ and $g^{(n)}$ are orthogonal w.r.t. the
Cartan-Killing form on $g$, i.e.
\begin{equation}
(g^{(m)},g^{(n)})=0
\end{equation}
iff $m \neq -n$.
\end{lemma}
Proof: Let $x \in g^{(m)}$ and $y \in g^{(n)}$, then
obviously
\be
([t_0,x],y)=m(x,y)
\ee
but also
\be
([t_0,x],y)=-(x,[t_0,y])=-n(x,y)
\ee
where we used the invariance property of the Cartan
Killing form. It follows immediately that
$(n+m)(x,y)=0$. Therefore if $n\neq -m$ we must have $(x,y)=0$. This
proves the lemma \vspace{7 mm}.

For notational convenience we shall sometimes denote the basis elements
$t_{j,m}^{(\mu)}$ simply by $t_a$ where $a$ is now the multi-index
$a\equiv (j,m;\mu )$. Let $K_{ab}$ denote the matrix components of
the Cartan-Killing form in this basis, i.e. $K_{ab}=(t_a, t_b)$
and let $K^{ab}$ denote its matrix inverse.

{}From the above lemma and the fact that the Cartan-Killing form is
non-degenerate on $g$ follows immediately that $g^{(k)}$ and $g^{(-k)}$
are  non-degenerately paired. This implies that if $t_a \in g^{(k)}$
then $K^{ab}t_b \in g^{(-k)}$ (where we used summation convention).

We then have the
following lemma which we shall need later.
\begin{lemma}
If $t_a$ is a highest weight vector (i.e. $a=(j,j;\mu )$ for some
$j$ and $\mu$) then $K^{ab}t_b$ is a lowest weight vector (and
vica versa). In particular if $t_a \in C(i)$ then $K^{ab}t_b \in C(i)$.
\end{lemma}
Proof: Since $t_a$ is  a highest weight vector we have
\begin{eqnarray}
0 & = & ([t_a,t_+],x) \nonumber \\
& = & (t_a,[t_+,x])
\end{eqnarray}
for all $x \in g$ which means that $t_a$ is orthogonal to
$\mbox{Im}(ad_{t_+})$ or put differently
\begin{equation}
\mbox{Ker}(ad_{t_+}) \perp \mbox{Im}(ad_{t_+})
\end{equation}
It is easy to do the same thing for $t_-$. One therefore has the
following decomposition of $g$ into mutually orthogonal spaces
\begin{equation}
g=\left( \mbox{Ker}(ad_{t_+}) + \mbox{Ker}(ad_{t_-}) \right) \oplus
\left( \mbox{Im}(ad_{t_+}) \cap \mbox{Im}(ad_{t_-}) \right) \nonumber
\end{equation}
where
\begin{equation}
\left( \mbox{Ker}(ad_{t_+}) + \mbox{Ker}(ad_{t_-}) \right) \perp
\left( \mbox{Im}(ad_{t_+}) \cap \mbox{Im}(ad_{t_-}) \right)
\end{equation}
Also one has the following decomposition
\begin{equation}
\mbox{Ker}(ad_{t_+})+\mbox{Ker}(ad_{t_-})=
C(i) \bigoplus_{k>0}\mbox{Ker}(ad_{t_+})
^{(k)} \bigoplus_{k>0} \mbox{Ker}(ad_{t_-})^{(-k)}
\end{equation}
As we saw above $K^{ab}t_b \in g^{(-k)}$ iff $t_a \in g^{(k)}$. Therefore
\begin{eqnarray}
\mbox{Ker}(ad_{t_{\pm}})^{(\pm k)} & \perp &
\mbox{Ker}(ad_{t_{\pm}})^{(\pm l)} \;\;\;\; \mbox{for
all }k,l>0 \nonumber \\
\mbox{Ker}(ad_{t_+})^{(k)} & \perp & \mbox{Ker}(ad_{t_-})^{(-l)} \;\;\;\;
\mbox{for } k \neq l \nonumber \\
\mbox{Ker}(ad_{t_+})^{(k)} & \perp & C(i) \;\;\;\;\;\;\; \mbox{for all }k>0
\nonumber \\
\mbox{Ker}(ad_{t_-})^{(-k)} & \perp & C(i)\;\;\;\;\;\;\; \mbox{for all }k>0
\nonumber
\end{eqnarray}
from which the lemma follows \vspace{7 mm}.

Note that from the proof of this lemma follows that the spaces
$\mbox{Ker}(ad_{t_+})^{(k)}$ and $\mbox{Ker}(ad_{t_-})^{(-k)}$
are nondegenerately
paired. The same is true for the centralizer $C(i)$ with itself.
We shall use these results in the next section when we start
reduction of the Kirillov Poisson structure.

\newsubsection{Reductions associated to $sl_2$ embeddings}

The procedure of Poisson reduction can be applied to Kirillov Poisson
structures to give new and in general nonlinear Poisson algebras.
If the Lie algebra is a KM algebra then, as was shown in \cite{BTV}
many $W$-algebras appearing in Conformal Field theory can be
constructed in this way. These Poisson reductions are associated
to inequivalent $sl_2$ embeddings into the finite underlying algebra
of the KM algebra. In \cite{tjark} it was announced that the same procedure
can be applied to finite dimensional Lie algebras and an interesting
special example was considered in detail. In this section the
general theory of these reductions will be developed.

Let there be given a certain $sl_2$ subalgebra $\{t_0,t_+,t_-\}$
of $g=sl_n$. Under the adjoint action of this $sl_2$ subalgebra
$g$ branches into a direct sum of irreducible $sl_2$ multiplets.
Let $\{t_{j,m}^{(\mu )}\}$ be the basis of $g$ introduced in the
previous section. Associated with this basis is a set of $C^{\infty}$
functions $\{ J^{j,m}_{(\mu)}\}$ on $g$ with the property
$J^{j,m}_{(\mu )}(t^{(\mu ' )}_{j',m'})=\delta^{\mu '}_{\mu}
\delta^j_{j'} \delta^m_{m'}$. These can be called the (global)
coordinate
functions on $g$ in the basis $\{t_{j,m}^{(\mu )}\}$
because they associate
to an element $\alpha \in g$ its $(j,m,\mu )$ component.

Let's now impose the following set of constraints
\be
\{\phi_{(\mu )}^{j,m}\equiv J^{j,m}_{(\mu )}-\delta_1^j\delta_1^m
\delta_{\mu}^1\}_{j \in \frac{1}{2}{\bf N}\; ; \; m>0} \label{constraints}
\ee
(remember that $t_{1,\pm 1}^{(1)} \equiv
t_{\pm}; \; t_{1,0}^{(1)} \equiv t_0$).
Denote the 'zero set'
in $g$ of these constraints by $g_c$. Its elements have the form
\begin{equation}
\alpha = t_+ + \sum_j \sum_{m\leq 0}\sum_{\mu} \alpha_{(\mu )}^{j,m}
t^{(\mu )}_{j,m}
\end{equation}
where $\alpha^{j,m}_{(\mu )}$ are real or complex numbers (depending
on which case we want to consider).

The constraints postulated above are motivated in the infinite
dimensional case by the requirement that the Poisson algebra which
we obtain after reduction must be a $W$ algebra \cite{BFFOW,BTV}.
In principle, from a mathematical point of view, one could consider
more general sets of constraints, however since we are primarily
interested in applications of our theory in conformal field theory
we shall restrict ourselves to the constraints (\ref{constraints}).

As discussed earlier we need to find out which of the constraints
(\ref{constraints})
are first class for they will generate gauge invariances on $g_c$. This
is the subject of the next lemma.
\begin{lemma}
The constraints $\{\phi_{(\mu )}^{j,m}\}_{m \geq 1}$ are first class.
\end{lemma}
Proof: First we show that
\begin{equation}
\{J^{j,m}_{(\mu )},J^{j',m'}_{(\mu ')}\}=\sum_{j'',\mu ''}
\alpha_{j''}^{\mu ''}J^{j'',m+m'}_{(\mu '')} \label{twee}
\end{equation}
for some coefficients $\alpha_{j}^{(\mu )}$. Let $x \in g^{(m'')}$,
then
\begin{equation}
\{J^{j,m}_{(\mu )},J^{j',m'}_{(\mu ')}\}(x)=
(x,[\mbox{grad}_xJ^{j,m}_{(\mu )},\mbox{grad}_xJ^{j',m'}_{(\mu ')}])
\label{een}
\end{equation}
Now let $y$ be an element of $g^{(k)}$, then
\begin{equation}
(y,\mbox{grad}_xJ^{j,m}_{(\mu )})=\frac{d}{d \epsilon}J^{j,m}_{(\mu )}
(x+\epsilon y)|_{\epsilon = 0}
\end{equation}
which is zero except when $k=m$ (see definition of $J^{j,m}_{(\mu )}$).
{}From this and the lemmas of the
previous section one concludes that $\mbox{grad}_xJ^{j,m}_{(\mu )}
\in g^{(-m)}$ and that the Poisson bracket (\ref{een}) is
nonzero if and only if $m''=m+m'$. From this eq.(\ref{twee}) follows.

Consider now the Poisson bracket between two constraints in the
set $\{\phi^{j,m}_{(\mu )}\}_{m \geq 1}$
\begin{eqnarray}
\{ \phi^{j,m}_{(\mu )},\phi^{j',m'}_{(\mu ')} \} & = &
\{J^{j,m}_{(\mu )},J^{j',m'}_{(\mu ')} \} \nonumber \\
& = & \sum_{j'',\mu ''}\alpha_{j'',\mu ''}\,J^{j'',m+m'}_{(\mu '')}
\nonumber \\
& = & \sum_{j'',\mu ''}\alpha_{j'',\mu ''}\,\phi^{j'',m+m'}_{(\mu '')}
\nonumber
\end{eqnarray}
which is obviously equal to zero on $g_c$
Note that the fact $m,m' \geq 1$ was used in the
last equality sign. This proves the lemma \vspace{7 mm}.

Note that in general the set $\{\phi_{(\mu )}^{j,m}\}_{m\geq 1}$
is not equal to the total set of constraints since the constraints
with $m=\frac{1}{2}$ are not included. These constraints will turn
out to be second class.

Let us now determine the group
of gauge transformations on $g_c$ generated by
the first class constraints. Again we use the multi-index notation
where now Roman letters $a,b,\ldots$ run over
all $j,m$ and $\mu$, and Greek letters $\alpha,\beta\ldots$
over all $j,\mu$ but only $m>0$ (those are the indices
associated with the constraints). Let $\phi^{\alpha}$
be one of the first class
constraints (i.e. $\alpha \equiv (j,m;\mu )$
with $m\geq 1$), then the gauge transformations associated to it are
generated by its Hamiltonian vectorfields
\begin{equation}
X_{\alpha}\equiv \{\phi^{\alpha},\cdot\}
\end{equation}
Let $x=x^at_a \in g$, then the change of $x$ under a gauge transformation
generated by $\phi^{\alpha}$ is given by
\begin{eqnarray}
\delta_{\alpha}x & = & \epsilon \{\phi^{\alpha},
J^{a}\}(x)t_a \nonumber \\
& = & \epsilon \{J^{\alpha},J^a\}(x)t_a \nonumber \\
& = & \epsilon g^{\alpha a}f^c_{ab}x^bt_c \nonumber \\
& = & [\epsilon g^{\alpha a}t_a,x]  \label{ga}
\end{eqnarray}
Since $g^{\alpha a} t_a \in g^{(-k)}$ iff $t_{\alpha}
\in g^{(k)}$ (see lemmas in the
previous section) we find that the Lie algebra of gauge transformations
is given by
\begin{equation}
h=\bigoplus_{k\geq 1}g^{(-k)}
\end{equation}
This is obviously a nilpotent Lie subalgebra
of $g$ and can be exponentiated to a
group $H$. This is the gauge group generated by the first class
constraints. Note that from eq.(\ref{ga}) follows that $H$ acts on
$g_c$ in the adjoint representation, i.e the gauge orbit of a point
$x \in g_c$ is given by
\begin{equation}
{\cal O} = \{ g x g^{-1} \mid g \in H \} \nonumber
\end{equation}

Now that we have identified the gauge group we can come to the matter
of constructing the space $g_c/H$, or equivalently, gauge fixing.
Of course this can be done in many ways, however in \cite{BFFOW,BTV}
it was argued that there are certain gauges which are the most convenient
from the conformal field theoretic point of view. These are
the so called 'lowest weight gauges'. Define the subset
$g_{fix}$ of $g_c$ as follows
\begin{equation}
g_{fix}=\{t_++\sum_{j,\mu} x^j_{(\mu )} t^{(\mu )}_{j,-j} \mid
x^j_{(\mu )}\in {\bf C}\}
\end{equation}
We then have the following
\begin{theorem}
\begin{equation}
H \times g_{fix} \simeq g_c
\end{equation}
\end{theorem}
Proof: Note first the obvious fact that $g^{(-l)}=g^{(-l)}_{lw} \oplus
g_0^{(-l)}$, where $g_0^{(-l+1)}=[t_+,g^{(-l)}]$ and $[t_-,g^{(-l)}_{lw}]
=0$. Let now $x \in g_c$ which means that $x=t_++ x^{(0)} + \ldots
+x^{(-p)}$ where $x^{(-k)} \in g^{(-k)}$ and $p \in \frac{1}{2} \bf N$
is the largest $j$ value in the decomposition of the adjoint
representation w.r.t. the embedding $i$.
Of course each $x^{(-k)}$ can
be written as a sum $x^{(-k)}_0 +x_{lw}^{(-k)}$ where $x^{(-k)} \in
g^{(-k)}_0$ and $x^{(-k)} \in g_{lw}^{(-k)}$.
Let also $\alpha^{(-k)}$ be an element of grade $-k$ in $h$, i.e.
$\alpha^{(-k)} \in g^{(-k)} \subset h$. Then
\begin{eqnarray}
\left( e^{\alpha^{(-k)}}x e^{-\alpha^{(-k)}} \right) ^{(-k+1)}
& = & (x^{(-k+1)}_0+x^{(-k+1)}_{lw})-ad_{t_+}(\alpha^{(-k)}) \nonumber \\
\left( e^{\alpha^{(-k)}}x e^{-\alpha^{(-k)}} \right) ^{(-k+i)}
& = & x^{(-k+i)} \;\;\;\;\;\; \mbox{for} \;\; 1 <i \leq k \nonumber
\end{eqnarray}
i.e. only the elements $x^{(-k+1)}, \ldots ,x^{(-p)}$ are changed by this
gauge transformation. Now, since $ad_{t_+}:g^{(-k)} \rightarrow
g^{(-k+1)}_0$ is bijective by definition, there exists a unique
$\alpha^{(-k)} \in g^{(-k)}$ such that $ad_{t_+}(\alpha^{(-k)})=
x_0^{(-k+1)}$, i.e. the element $x^{(-k+1)}_0$ can be gauged away by
choosing $\alpha^{(-k)}$ appropriately.
{}From this follows immediately that there exist unique
elements $\alpha^{(-l)}\in g^{(-l)}$ ($l=1, \ldots ,p$) such that
\begin{equation}
e^{\alpha^{(-p)}}\ldots e^{\alpha^{(-1)}}x \,e^{-\alpha^{(-1)}} \ldots
e^{-\alpha^{(-p)}}= y \;\;
\in g_{fix}
\end{equation}
This provides, as
one can now easily see, a
bijective map between $H \times g_{fix}$ and $g_c$ given by
\begin{equation}
\left( e^{-\alpha^{(-1)}} \ldots e^{-\alpha^{(-p)}}, y \right)
\rightarrow x
\end{equation}
This proves the theorem \vspace{7 mm}.

So starting from $g$, after imposing constraints and fixing gauge
invariances we have arrived at a submanifold $g_{fix}$ of $g$.
We now want to determine the Poisson algebra structure of $C^{\infty}
(g_{fix})$. For this we need to calculate the Dirac brackets
\begin{equation}
\{J^{j,-j}_{(\mu )},J^{j',-j'}_{(\mu ')}\}^*
\end{equation}
between the generators $\{J^{j,-j}_{(\mu )}\}$ of $C^{\infty}(g_{fix})$.
We shall first address a slightly more general problem and then
specialize to reductions associated to $sl_2$ embeddings.

Let $\{t_i\}_{i=1}^{dim(g)}$ be a basis of the Lie algebra $g$, let
$k$ be a positive integer smaller or equal to $\mbox{dim}(g)$ and
denote $t_{k+1} \equiv \Lambda$. Consider then the following subset
of $g$
\begin{equation}
g_f=\{\Lambda +\sum_{i=1}^{k}\alpha^i t_i \mid \alpha^i \in {\bf C} \}
\end{equation}
which can be seen as the zero set of the constraints $\phi^1=J^{k+1}-1$
and $\phi^i=J^{k+i}$ for $1<i\leq p-k$. Also suppose that the Kirillov
bracket on $g$ induces a Dirac bracket
$\{.,.\}^*$ on $g_f$ (i.e. all constraints
are second class).

Denote by $\cal R$ the set
of smooth functions
\begin{equation}
R:{\bf C}^k \times g_f \longrightarrow g
\end{equation}
of the form
\begin{equation}
R(\vec{z};y)=\sum_{i=1}^{k} z_iR^i(y)
\end{equation}
where $\vec{z}\equiv \{z_i\}_{i=1}^k \in {\bf C}^k$, $y \in g_f$ and
$R^i(y) \in sl_n$. To any element $R \in {\cal R}$ one can associate
a map
\begin{equation}
Q_R : {\bf C}^k \times g_f \longrightarrow C^{\infty}(g)
\end{equation}
defined by
\begin{equation}
Q_R(\vec{z};y)=\sum_{i=1}^{dim(g)} (R(\vec{z};y),t_i)J^i
\end{equation}
(where as before $J^i \in C^{\infty}(g)$ is defined by $J^i(t_j)=
\delta_j^i$). We are going to use certain elements of the set $\cal R$
in order to explicitly calculate the Dirac brackets on $g_f$. We have
the following theorem.
\begin{theorem}
If there exists a unique $R \in \cal R$ such that for all $\vec{z} \in
{\bf C}^k$ and $y \in g_f$ we have
\be
\Lambda +[R(\vec{z};y),y] \;\;\;\; \in g_{f} \label{r1}
\ee
and
\be
Q_R(\vec{z};y)|_{g_{f}}=\sum_{i=1}^k z_iJ^i
+\left( R(\vec{z};y),\Lambda \right)   \label{r2}
\ee
(i.e. the restriction of $Q_R(\vec{z};y)$ to $g_f$ is equal to the
right hand side of (\ref{r2})) then
\begin{equation}
\sum_{ij=1}^k z_i\; \{J^i,J^j\}^*(y)
\; t_j=[R(\vec{z};y),y] \label{trick}
\end{equation}
for all $\vec{z} \in {\bf C}^{k}$ and $y \in g_f$.
\end{theorem}
Note that from eq.(\ref{trick}) one can read off all the Dirac brackets
between the generators $\{J^i\}_{i=1}^k$ of $C^{\infty}(g_f)$ since
the formula holds for all $\vec{z} \in {\bf C}^k$ and the elements
$t_j$ are all independent. The only thing that one therefore needs
to determine is the map $R$.  Also note that from equation (\ref{r2})
follows that within the Dirac bracket $\{.,.\}^*$ the function
$Q_R(\vec{z};y)$ is equal to $\sum_{i=1}^kz_iJ^i$, i.e.
\be
\{Q_R(\vec{z};y),\cdot \}^* =\sum_{i=1}^{k} z_i\{J^i,\cdot \}^*
\ee
since constants commute with everything and restriction to $g_f$
is always implied within the Dirac bracket.

The proof of the theorem is based on the following two lemmas.
\begin{lemma}
Let $M$ be a Poisson manifold, $Q \in C^{\infty}(M)$ and $X_Q$
its Hamiltonian vectorfield, i.e. $X_Q=\{Q,.\}$. Let there also be
given a set of second class constraints and let $Z$ be the submanifold
where they all vanish. If for some $p \in M$ we have $X_Q|_p(\phi_i)=0$
for all $i$ (i.e. $X_Q|_p$ is tangent to $Z$), then
\begin{equation}
\{Q,F\}(p)=\{\bar{Q},\bar{F}\}^*(p) \;\; \;\; \mbox{for all} \;\;F \in
C^{\infty}(M)
\end{equation}
where $\{.,.\}^*$ is the Dirac bracket.
\end{lemma}
Proof: Simply calculating the Dirac bracket in $p$ gives
\begin{eqnarray}
\{\bar{Q},\bar{F}\}^*(p) & \equiv & \{Q,F\}(p)-\{Q,\phi_i\}(p)
\Delta^{ij}(p)
\{\phi_j,F\}(p)  \nonumber \\
& = & \{Q,F\}(p) \nonumber
\end{eqnarray}
because by hypothesis $\{Q,\phi_i\}(p)=0$ for all $i$.
This proves the lemma \vspace{4 mm}.

The origin of equation (\ref{r1}) in the theorem becomes clear
in the following lemma.
\begin{lemma}
If for all $y \in g_f$ and $\vec{z} \in {\bf C}^k$
\be
\Lambda +[R(\vec{z};y),y] \;\;\;\; \in g_f
\ee
then
\be
\{Q_R(\vec{z};y),\phi^i\}(y)=0
\ee
for all $1 \leq i \leq p-k;$ $y \in g_f$ and $\vec{z} \in {\bf C}^k$.
\end{lemma}
Proof: By definition of $J^i$ we have $y=t_iJ^i(y)$ which
gives us
\begin{eqnarray}
\Lambda +[R(\vec{z};y),y] & = & \Lambda +
\sum_{ij=1}^{dim(g)} [R(\vec{z};y)^it_i,t_jJ^j](y) \nonumber \\
& = & \Lambda +
\sum_{ijlm=1}^{dim(g)} K_{ij}f^{jl}_{m}R
(\vec{z};y)^iJ^m(y)t_l \nonumber \\
& = & \Lambda +\sum_{ijl=1}^{dim(g)}\{K_{ij}R(\vec{z};y)^iJ^j,J^l\}(y)t_l
\nonumber \\
& \equiv & \Lambda +\sum_{i=1}^{dim(g)} \{Q_R(\vec{z};y),J^i\}(y)t_i
\nonumber
\end{eqnarray}
but by hypothesis this was an element of $g_{f}$. Therefore
$\{Q_R(\vec{z};y),J^i\}(y)=0$ for $i \geq k+1$.
This proves the lemma \vspace{4 mm}.

Putting together these two lemmas we find

Proof of theorem: From the previous lemmas follows that if eqns.
(\ref{r1}) and (\ref{r2})   are satisfied, then
\begin{eqnarray}
\sum_{ij=1}^{k}z_i \{J^i,J^j\}^*(y)t_j
& = & \sum_{i=1}^k \{Q_R(\vec{z};y),J^i\}^*(y)t_i
\nonumber \\
& = & \sum_{i=1}^k \{Q_R(\vec{z};y),J^i\}(y)t_i \nonumber \\
& = & \sum_{i=1}^{dim(g)}\{Q_R(\vec{z};y),J^i\}(y)t_i \nonumber \\
& = & [R(\vec{z};y),y]
\end{eqnarray}
where eqn.(\ref{r2}) was used in the first step.
This proves the theorem \vspace{7 mm}.

Note that conversely in order to show the
existence of the Dirac bracket on $g_{f}$
it is sufficient to prove that equations (\ref{r1}) and (\ref{r2}) of the
above theorem are solvable within $\cal R$.
We will now show that
this is the case when $g_f \equiv g_{fix}$ associated to an arbitrary
$sl_2$ embedding.
\begin{theorem}
Let $|i|$ be the total number of $sl_2$ multiplets in the
branching of the adjoint representation of $g$. The
equations
\begin{eqnarray}
&   & t_++[R(\vec{z};y),y] \;\;\;\; \in g_{fix}\;\;\;\;\; \mbox{for all }
\vec{z} \in
{\bf C}^{|i|} \mbox{ and } y \in g_{fix} \nonumber \\
&   & Q_R(\vec{z};y)|_{g_{fix}}= \sum_{j;\mu}
z^{(\mu )}_j J^{j,-j}_{(\mu)}+\left( R(\vec{z};y),t_+ \right)
\;\;\;\;\mbox{for
all } z \in {\bf C}^{|i|} \mbox{ and } y \in g_{fix}\nonumber
\end{eqnarray}
have a unique solution $R\in \cal R$ for
arbitrary embeddings.
\end{theorem}
Proof: First solve the second equation.
\begin{eqnarray}
Q_R(\vec{z};y)(y) & = & \left( R^{(p)}+ \ldots +R^{(-p)},
t_++y^{(0)}+\ldots
+y^{(-p)} \right) \nonumber \\
& = & (R^{(-1)},t_+)+(R^{(p)},y^{(-p)}) +\ldots +(R^{(0)},
y^{(0)}) \nonumber
\end{eqnarray}
Now, $y^{(-j)}\in g^{(-j)}_{lw}$ which means that $y^{(-j)}=\sum_{\mu }
y^{j,-j}_{(\mu )}t_{j,-j}^{(\mu )}$. Also $R^{(j)}=R^{(j)}_0+R^{(j)}
_{hw}$ where $R^{(j)}_0 \in \mbox{Im}(ad_{t_+})$ and $R^{(j)}_{hw} \in
\mbox{Ker}(ad_{t_+})$. Let $\{t^{j,-j}_{(\mu )}\}_{\mu } \subset g^{(j)}$
be such that $(t_{j,-j}^{(\mu )},t^{j',-j'}_{(\mu ')})=\delta^{j'}_j
\delta^{\mu }_{\mu '}$ (remember that $g^{(j)}$ and $g^{(-j)}$ are
non-degenerately paired). Then take
\begin{equation}
R_{hw}^{(j)}=\sum_{\mu} z_j^{(\mu)}t_{(\mu)}^{j,-j}
\end{equation}
Obviously the second equation of the theorem is then satisfied.

Let's now solve the first equation. For $k>0$ it reads
$[R,y]^{(k)}=0$. Writing this out gives
\begin{equation}
[R^{(k-1)}_0,t_+]=[y^{(0)},R^{(k)}]+ \ldots + [y^{(k-p)},R^{(p)}]
\label{342}
\end{equation}
Remember that $ad_{t_+}:g^{(k-1)}_0 \rightarrow g^{(k)}$ is bijective
which means that if we already know  $R^{(k)}, \ldots ,R^{(p)}$ then we
can uniquely solve this equation for $R_0^{(k-1)}$ (remember that
for $l\geq 0$ we have already determined the quantities $R_{hw}^{(l)}$).
The initial term of this sequence of equations is
obviously $R^{(-p)}=R^{(-p)}_{
hw}$ (i.e. $R^{(-p)}_0=0$). In this way we can completely determine
the positive and zero grade piece of $R(\vec{z};y)$.

For $-k \geq 0$ the situation is similar but slightly different. What
we now have to demand is that $[R,y]^{(-k)}_0=0$. This leads
to the equation
\begin{equation}
[t_+,R^{(-k-1)}]=[R^{(-k)},y^{(0)}]_0+ \ldots + [R^{(-k+p)},y^{(-p)}]_0
\label{343}
\end{equation}
where we used that $[t_+,R^{(-k-1)}]$ is already an element of
$g^{(-k)}_0$. Remember that $ad_{t_+}:g^{(-k-1)} \rightarrow
g_0^{(-k)}$ is bijective so there exists a unique $R^{(-k-1)}(z)$
such that the equation is solved, assuming we know $R^{(-k)},\ldots ,
R^{(-k+p)}$ already. In this way we can grade  by grade determine the
negative grade piece of the matrix $R$.
It is also clear
from the above construction that
$R \in \cal R$. This proves the theorem \vspace{7 mm}.

It is possible to derive a general formule for $R(\vec{z};y)$ using
arguments similar to those used in \cite{BG}. Let again
\begin{equation}
g_{lw}= \bigoplus_{j \in \frac{1}{2} \bf N} g^{(-j)}_{lw}=\mbox{span}(
\{t^{(\mu )}_{j,-j}\}_{j;\mu }
\end{equation}
and let $\Pi$ be the orthogonal (w.r.t. the Cartan-Killing
form) projection onto $\mbox{Im}(ad_{t_+})$. Obviously the map
$ad_{t_+}: \mbox{Im}(ad_{t_-}) \rightarrow \mbox{Im}(ad_{t_+})$ is
invertible. Denote the inverse  of this map, extended by $0$ to
$g$ by $L$. As before what we want to do is solve the equation
\begin{equation}
[R,y]=x \; \in g_{lw}
\end{equation}
for $y \in g_{fix}$. Noting that $y=t_++w$ where $w \in g_{lw}$
and applying $\Pi$ this equation reduces to
\begin{equation}
\Pi \circ ad_{t_+}R=\epsilon \Pi ([R,w])
\label{master}
\end{equation}
where we introduced a parameter $\epsilon$ in the right hand side
which we want to put to 1 later. Note that the left hand side
is equal to $ad_{t_+}R$ since this is already an element of
$\mbox{Im}(ad_{t_+})$. Assume now that $R$ can be (perturbatively)
written as
\begin{equation}
R=\sum_{k=0}^{\infty} R_k\epsilon^k
\end{equation}
(we shall have to justify this later). Consider now the zeroth order
part of the equation (\ref{master}). It reads
\begin{equation}
ad_{t_+}R_{0}=0
\end{equation}
This means that $R_0\equiv F(\vec{z})$
which is an arbitrary element of $\mbox{ker}(ad_{t_+})$.
The first order equation is equal to
\begin{equation}
ad_{t_+}R_1=\Pi ([F,w])
\end{equation}
Obviously this equation is solved by
\begin{equation}
R_1=-L\circ ad_wF
\end{equation}
Proceeding with higher orders we find
\begin{equation}
R_{k+1}(\vec{z};y)=-L\circ ad_w \left( R_k(\vec{z};y) \right)
\end{equation}
which means that
\begin{equation}
R(\vec{z};y)=\frac{1}{1+\epsilon L\circ ad_w}F(\vec{z})
\end{equation}
There are no convergence problems with this series since the
operator $L$ lowers the degree by one which means that after $2p$
steps it must cancel. We therefore find
\begin{equation}
\sum_{j,j';\mu \mu '}z^{(\mu )}_j \; \{J^{j,-j}_{(\mu )},
J^{j',-j'}_{(\mu ')}\}^*(y)\; t_{j',-j'}^{(\mu ')}=
-ad_y \left( \frac{1}{1+L ad_w}F(\vec{z}) \right)
\end{equation}
Let again $y=t_++w \in g_{fix}$ (i.e. $w \in g_{lw}$) and $Q \in
C^{\infty}(g_{fix})$. We then define $\mbox{grad}_yQ \in \mbox{Ker}(
ad_{t_+}) \equiv g_{hw}$ by
\begin{equation}
\left( x,\mbox{grad}_y Q \right) \equiv \frac{d}{d\epsilon}
Q(y +\epsilon x)|_{\epsilon=0} \;\;\;\mbox{for all }x \in g_{lw}
\end{equation}
Note that this uniquely defines $\mbox{grad}_yQ$ because, as we saw
before, $g_{lw}$ and $g_{hw}$ are non-degenerately paired. It
is now easy to derive a general (and coordinate free) formula for the
(Dirac) Poisson structure on $C^{\infty}(g_{fix})$ induced by the
Kirillov Poisson structure on $C^{\infty}(g)$. For $Q_1,Q_2 \in
C^{\infty}(g_{fix})$ it reads
\begin{equation}
\{Q_1,Q_2\}^*(y) =\left(y,[ \mbox{grad}_yQ_1, \frac{1}{1+L
\circ ad_w}\mbox{grad}_yQ_2] \right) \label{PA}
\end{equation}
For the so called trivial embedding ($t_0=t_{\pm}=0$)
the map $L$ is equal to the zero map and the above formula reduces
to  the ordinary Kirillov bracket as it should (because in that
case $g_{fix} \equiv g$).

{}From now on denote the Kirillov Poisson algebra associated to a
(semi) simple Lie algebra $g$ by
$K(g)$ and the Poisson algebra $C^{\infty}
(g_{fix})$ with Poisson structure (\ref{PA}) by ${\cal W}(i)$ where
$i$ is again the $sl_2$ embedding in question.

In general ${\cal W}(i)$ is a non-linear Poisson algebra
as we shall see when we consider examples. However it can happen that
${\cal W}(i)$ contains a subalgebra that is isomorphic to a Kirillov
algebra. The next theorem deals with this.
\begin{theorem}
The algebra ${\cal W}(i)$ has a Poisson subalgebra which is
isomorphic to
\begin{equation}
\bigotimes_{j \in \frac{1}{2} \bf N} K(sl_{n_j}) \bigotimes (q-1)K(u(1))
\label{sub}
\end{equation}
(where again $q$ is the
number of different $j$ values in the decomposition of the
fundamental representation).
\end{theorem}
Proof: Remember that the centralizer of the $sl_2$ subalgebra $i(sl_2)$
is given by
\begin{equation}
C(i)=\bigoplus_{j \in \frac{1}{2} \bf N} sl_{n_j} \bigoplus (q-1)u(1)
\end{equation}
and therefore the Kirillov algebra of $C(i)$ is equal to (\ref{sub}).
Since all elements of the centralizer are lowest
(and highest) weight vectors w.r.t.
$i(sl_2)$ the function $J^a$ is not constrained if $t_a \in C(i)$,
i.e. all the elements $J^a$ associated to the centralizer survive the
reduction. It is not obvious however that they still form the algebra
(\ref{sub}) w.r.t. the Dirac bracket. This is what we have to show.

The part of the equation
\begin{equation}
\sum_{j,j';\mu ,\mu '}z_j^{(\mu )} \{J^{j,-j}_{(\mu )},J^{j',-j'}_{(\mu ')}
\}^*(y)t_{j',-j'}^{(\mu ')}=[R(\vec{z};y),y]
\end{equation}
that determines the Poisson relations
between the currents associated to the centralizer is
\begin{equation}
\sum_{\mu,\mu '}z^{(\mu )}_0 \; \{J^{0,0}_{(\mu )},J^{0,0}_{(\mu ')}\}^*
(y) \; t_{0,0}^{(\mu ')}=[R(\vec{z};y),y]^{(0)}
\end{equation}
The right hand side of this equation reads in more detail
\begin{equation}
[R,y]^{(0)}=[R^{(0)}_{lw},y^{(0)}]+[R^{(0)}_0,y^{(0)}]
+ \ldots + [R^{(p)},y^{(-p)}]+
[R^{(-1)},t_+]   \label{tja}
\end{equation}
Note that $R^{(0)}_{lw}$ and $y^{(0)}$ are both elements of $C(i)$
which means that the first term in the right hand side is also
an element of $C(i)$. We will now show that $R^{(0)}_0,R^{(1)}, \ldots,
R^{(p)}$ and $R^{(-1)}$ do not depend on $\{z^{(\mu )}_0 \}_{\mu}$
which means that all but the first term in the right hand side of
equation (\ref{tja}) are irrelevant for the Poisson brackets
$\{J^{0,0}_{(\mu )},J^{0,0}_{(\mu '}\}^*$.

{}From eq.(\ref{342}) follows that $R^{(k)}_0$ is only a function of
$z^{(\mu )}_p, \ldots ,z^{(\mu )}_{k+1}$ for $k\geq 0$ and therefore
$R^{(k)}=R^{(k)}(z^{(\mu )}_p, \ldots ,z^{(\mu )}_k)$. Now, $R^{(-1)}$
is determined by equation (\ref{343}) for $k=0$
\begin{equation}
[t_+,R^{(-1)}]=[R^{(0)},y^{(0)}]_0+ \ldots +[R^{(p)},y^{(-p)}]_0
\end{equation}
Note that the terms $[R^{(l)},y^{(-l)}]$ for $l>0$
certainly do not contain $z_0^{(\mu )}$ as we just saw. Also
note that
\begin{equation}
[R^{(0)},y^{(0)}]_0=[R^{(0)}_0+R^{(0)}_{lw},y^{(0)}]_0=
[R^{(0)}_0,y^{(0)}]
\end{equation}
because of the reason we mentioned earlier that $R^{(0)}_{hw}$
and $y^{(0)}$ are both in $C(i)$. However, as we have seen above
$R^{(0)}_0$ does not contain $z^{(\mu )}_0$. From this we conclude
that the only term in $[R,y]^{(0)}$ that contains $z_0^{(\mu )}$
is $[R^{(0)}_{lw},y^{(0)}]$, i.e.
\begin{eqnarray}
\sum_{\mu \mu '}z^{(\mu )}_0 \{J^{0,0}_{(\mu )},J^{0,0}_{(\mu ')}\}
^* (y) t_{0,0}^{(\mu )} & = & [R^{(0)}_{lw},y^{(0)}] \nonumber \\
& = & [R_{lw}^{(0)}, \sum_{\mu } J^{0,0}_{(\mu )}(y)t_{0,0}^{(\mu )}]
\label{hup3}
\end{eqnarray}
As the generators $\{t_{0,0}^{(\mu )}\}$ are a basis of $C(i)$ they
form a
\begin{equation}
\bigoplus_{j} sl_{n_j} \bigoplus (q-1)u(1)
\end{equation}
algebra. From this, eq.(\ref{hup3}) and theorem 2 follows
immediately that the Poisson algebra
generated by $\{J^{0,0}_{(\mu )}\}$ w.r.t. the Dirac
bracket is isomorphic to  Kirillov
algebra of $C(i)$. This proves the theorem \vspace{7mm}.

\newsubsection{Generalized finite Miura transformations}

In this section we will present a generalized version of the Miura
transformation. Roughly this is a Poisson homomorphism of the
finite $W$ algebra ${\cal W}(i)$
in question to a certain Kirillov algebra.
In order to be able to describe this map
for arbitrary embeddings however we first have
to concern ourselves with the cases when in the decomposition of $g$
into $sl_2$ multiplets there appear half integer grades. As we have seen, in
those cases the constraints $\phi^{j,\frac{1}{2}}_{(\mu )}$ are
second class. In what follows it will be necessary to be able
to replace the usual set of constraints by an alternative set which
contains only first class constraints but which gives rise to the
same reduction \cite{FORTW}. Roughly what one does is impose only half of the
constraints that turned out to be second class in such a way that
they become first class. The other constraints that were second class
can then be obtained as gauge fixing conditions. In this way $g_{fix}$
stays the same but $g_c$ is different. Since the resulting Poisson
algebra only depends on $g_{fix}$ it is clear that we obtain the same
algebra ${\cal W}(i)$.

Let's now make all of this more precise.
We describe the $sl_n$ algebra in the standard way by traceless
$n\times n$ matrices; $E_{ij}$ denotes the matrix with a one in
its $(i,j)$ entry and zeroes everywhere else. As we said earlier
embeddings of $sl_2$ into $sl_n$ are in
one-to-one correspondence with partitions of $n$. Let $(n_1,n_2,\ldots)$ be
a partition of $n$, with $n_1\geq n_2 \geq \ldots$. Define a different
partition $(m_1,m_2,\ldots)$ of $n$, with $m_k$ equal to the
number of $i$ for which $n_i\geq k$. Furthermore let
$s_t=\sum_{i=1}^{t} m_i$. Then we have the following
\bl
An embedding of $sl_2$ in $sl_n$ under which the fundamental
representation branches according to $n\rightarrow \oplus n_i$
is given by
\ba
t_+ & = & \sum_{l\geq 1} \sum_{k=1}^{n_l-1} E_{l+s_{k-1},l+s_k},
\nonu
t_0 & = & \sum_{l\geq 1} \sum_{k=1}^{n_l} (\frac{n_l+1}{2}-k)
E_{l+s_{k-1},l+s_{k-1}}, \nonu
t_- & = & \sum_{l\geq 1} \sum_{k=1}^{n_l-1}  k
(n_l-k)
E_{l+s_k,l+s_{k-1}}.
\label{embedding}
\ea
\el
The proof is by direct computation\footnote{The commutation
relations are $[t_0,t_{\pm}]=\pm t_{\pm}$, and $[t_+,t_-]=
2t_0$.}. If the fundamental
representation of $sl_n$ is spanned by  vectors $v_1,\ldots v_n$,
on which $sl_n$ acts via $E_{ij}(v_k)=\delta_{jk}v_i$,
then the irreducible representations $n_l$ of $sl_2$
to which it reduces are spanned by $\{v_{l+s_{k-1}}\}_{1\leq k
\leq n_l}$. Previously we decomposed the lie algebra $g$ into
eigenspaces of $ad_{t_0}$. However, for our present purposes it is
convenient to introduce a different grading of the Lie algebra
$g$, that we need
to describe the generalized Miura map and
the BRST quantization. The grading is
defined by the following element of the Cartan subalgebra of
$sl_n$:
\be
\label{def:delta}
\delta=\sum_{k\geq 1} \sum_{j=1}^{m_k} \left( \frac{\sum_l
lm_l}{\sum_l m_l} -k \right) E_{s_{k-1}+j,s_{k-1}+j}.
\ee
This leads to the alternative decomposition $g=g_-\oplus g_0 \oplus g_+$ of
$g$ into spaces with negative, zero and positive eigenvalues under
the adjoint action of $\delta$ respectively (note that in
case the grading of $ad_{t_0}$ is an integral grading then
we have $t_0=\delta$ so in those cases nothing happens. In general
however $t_0\neq \delta$ and also $g^{(m)}\neq g_m$).  The subalgebra
$g_0$ consists of matrices whose nonzero entries are in square
blocks of dimensions $m_1\times m_1$, $m_2\times m_2$, etc.
along the diagonal of the matrix and is therefore
a direct sum of $sl_{m_k}$ subalgebras (modulo $u(1)$ terms).
The nilpotent subalgebra $g_+$
is spanned by
$\{E_{l+s_{k-1},r+s_k}\}_{l\geq 1;\; 1\leq k \leq
n_l-1;\; r>0}$,
and the nilpotent subalgebra $g_-$ by the transpose
of these. Let $\pi_{\pm}$ denote the projections onto $g_{\pm}$.
Then  the following theorem describes the replacement of the mixed
system of first and second class constraints by a system of first class
constraints only.
\bt
\label{firstclass}
The constraints
$\{J^{l+s_{k-1},r+s_k}-\delta^{r,l} \}_{l\geq 1;\; 1\leq k \leq
n_l-1;\; r>0}$
are first class. The gauge group
they generate is $\hat{H}=\exp(g_-)$ acting via the adjoint
representation on $g$. The resulting finite $W$ algebra is the
same as the one obtained by imposing the constraints (\ref{constraints}).
\et
Proof: decompose $g$ in eigenvalues of $\ad{\delta}$,
$g=\oplus_k g_{k}$. Note that $\ad{\delta}$ has only
integral eigenvalues. Using the explicit form of $t_+$ in
(\ref{embedding}), one easily verifies that $[\delta,t_+]=t_+$.
Thus, $t_+\in g_{1}$. Again since $[g_+,g_+]=[g_{\geq
1},g_{\geq 1}]\subset g_{\geq 2}$ it follows
(exactly like in lemma 3)  that the constraints
$\{J^{l+s_{k-1},r+s_k}-\delta^{r,l} \}_{l\geq 1;\; 1\leq k \leq
n_l-1;\; r>0}$
are first class.
The gauge group  can be determined similarly as in section 3.2,
and the analogue of theorem 1 of section 3.2 can be proven in the
same way with the same choice of $g_{fix}$, if one uses the
decomposition of $g$ in eigenspaces of $\ad{\delta}$ rather than
$\ad{t_0}$. Therefore the resulting finite $W$ algebra is the
same, because theorem 2 and 3 of section 3.2 show that it only
depends on the form of $g_{fix}$.\\

Let us explain this theorem in words. The number of second
class constraints in any system is necessarily even. If one switches
to the $ad_{\delta}$ grading the set of second class constraints is
split into half:
one half gets grade 1 w.r.t. $ad_{\delta}$
while the other half gets grade 0. Now what one does is impose only that
half that has obtained grade 1 w.r.t. the $ad_{\delta}$ grading. These
constraints are then first class. The gauge transformations they
generate can be completely fixed by imposing the constraints that
were in the other half. Having done that we are back in exactly the
same situation as before. The only difference is that we now know of
a system of first class constraints that in the end leads to the
same reduction.

Note that the number of generators of the finite $W$ algebra is equal to
$\dim(g_0)=(\sum_i m_i^2) -1$. This is indeed the same as the
number of irreducible representations of $sl_2$, minus one, one
obtains from $(\oplus_i n_i)\otimes (\oplus_i n_i)$, as the
latter number equals $(\sum_j (2j-1)n_j)-1$, and one easily
checks that $\sum_i m_i^2=\sum_j (2j-1)n_j$.

The generalized Miura transformation can now be formulated as
\begin{theorem}
There exists an injective Poisson homomorphism from ${\cal W}(i)$ to
$K(g_0)$.
\end{theorem}
Proof: First we show that
for every element $x \in t_++g_0$ there exist a unique element
$h$ in the gauge group $\hat{H}$ such that $h\cdot x \cdot h^{-1} \in
g_{fix}$.
For this note that there exists
a unique element $h' \in \hat{H}$ such that $h'.x.h'^{-1} \in g_c$.
This follows from the previous theorem and the remarks made after it.
In fact it follows from those remarks that $h'$ is an element of
that subgroup of $\hat{H}$ that is generated by the 'ex' second
class constraints (the ones that were made into first class constraints
by not imposing the other half). It was shown earlier however that
\be
g_{fix}\times H  \simeq g_c
\ee
which means that there exists a unique element $h'' \in H$ such that
\be
h''h'.x.(h''h')^{-1} \in g_{fix}   \label{nulfix}
\ee
We conclude from this that there is a surjective map from $x \in t_++g_0$
to $g_{fix}$ given by (\ref{nulfix}).
The pull back of this map $C^{\infty}(g_{fix})\equiv {\cal W}(i)
\rightarrow C^{\infty}(t_++g_0)$ (which will therefore
be injective) is then the Miura map. Of course we still have to check
whether this map is a Poisson homomorphism. This we address next.

What is the Poisson structure on $C^{\infty}(t_++g_0)$ ?
Since the constraints
of which the space $t_++g_0$ is the zero set are obviously second
class the Kirillov bracket on $g$ induces a Dirac bracket on
it. It is not difficult to see that the Dirac term in the Dirac
bracket cancels in this case which means that the Poisson algebra
$C^{\infty}(t_++g_0)$ with the induced Poisson structure is
isomorphic to the Poisson algebra $K(g_0)$. Since the transformation
from ${\cal W}(i)$ to $K(g_0)$ corresponds to a gauge transformation
(\ref{nulfix}) the map is necessarily a homomorphism.

\newsubsection{Examples}

The simplest examples of finite $W$ algebras are those associated
to the so called 'principal $sl_2$ embeddings'.
These embeddings are associated to the trivial partition of the
number $n$: $n=n$.
The fundamental representation of $g=sl_n$ therefore becomes
an irreducible representation of the $sl_2$ subalgebra, i.e.
\be
\underline{n}_n \simeq \underline{n}_2
\ee
The branching rule for the adjoint representation of $g$ therefore
reads
\be
\underline{ad}_n \simeq \underline{3}_2 \oplus \underline{5}_2\oplus
\ldots \oplus \underline{2n-1}_2
\ee
{}From this follows immediately that the finite $W$ algebra will have
$n-1$ generators (since there are $n-1$ $sl_2$ multiplets). Without
going into details we can immediately predict the Poisson relations
between these generators from the generalized Miura transformation
for in this case the subalgebra $g^{(0)}=g_0$ coincides with the
Cartan subalgebra of $sl_n$. Since the Cartan subalgebra is an
abelian algebra, and since the Kirillov algebra of an abelian Lie
algebra is a Poisson commutative algebra we find that a finite $W$
algebra associated to a principal embedding
must also be Poisson commutative since it is isomorphic to a Poisson
subalgebra of
$K(g_0)$. We conclude therefore that the principal $sl_2$ embedding
into $sl_n$ leads to the abelian Poisson algebras with $(n-1)$ generators
\vspace{3mm}.

The simplest nontrivial case of a finite $W$ algebra is associated to
the (only) nonprincipal embedding of $sl_2$ into $sl_3$. This embedding
is associated to the following partition of 3: $3 \rightarrow 2+1$.
The branching rule of the fundamental representation of $sl_3$ is
therefore
\be
\underline{3}_3 \simeq \underline{2}_2 \oplus \underline{1}_2
\ee
{}From this we find the following branching rule for the adjoint
representation
\be
\underline{ad}_3 \simeq \underline{3}_2 \oplus 2.\underline{2}_2 \oplus
\underline{1}_2
\ee
from which follows immediately that the finite $W$ algebra associated
to this embedding will have 4 generators. We shall go through the
construction of this finite $W$ algebra in some detail in order to illustrate
the theory discussed above.

The explicit form of the $sl_2$ embedding is
\begin{eqnarray}
t_+ & = & E_{1,3} \nonumber \\
t_0 & = & \mbox{diag}(\frac{1}{2},0,-\frac{1}{2}) \nonumber \\
t_- & = & E_{3,1}
\end{eqnarray}
where as before $E_{ij}$ denotes the matrix with a one in its $(i,j)$
entry and zeros everywhere else. The ($sl_3$ valued) function
$J=t_{j,m}^{(\mu )}J^{j,m}_{(\mu )}$ (where we used summation convention)
reads
\be
J=
\left(
\begin{array}{ccc}
\frac{1}{2}J^{1,0}_{(1)}+J^{0,0}_{(1)} & J^{\frac{1}{2},\frac{1}{2}}_{(1)}
& J^{1,1}_{(1)} \\
J^{\frac{1}{2},-\frac{1}{2}}_{(2)} & -2J^{0,0}_{(1)} & J^{\frac{1}{2},
\frac{1}{2}}_{(2)} \\
J^{1,-1}_{(1)} & J^{\frac{1}{2},-\frac{1}{2}}_{(1)} & J^{0,0}_{(1)}
-\frac{1}{2}J^{1,0}_{(1)}
\end{array} \right)
\ee
According to the general prescription the constraints are
\be
J^{1,1}_{(1)}-1=J^{\frac{1}{2},\frac{1}{2}}_{(1)}
=J^{\frac{1}{2},\frac{1}{2}}_{(2)}=0
\ee
the first one being the only first class constraint. As was shown
earlier the gauge invariance generated by this constraint can be
completely fixed by adding the 'gauge fixing condition'
\be
J^{1,0}_{(1)}=0
\ee
The Dirac brackets between the generators $\{J^{0,0}_{(1)},
J^{\frac{1}{2},-\frac{1}{2}}_{(1)},J^{\frac{1}{2},-\frac{1}{2}}_{(2)}\}$
and $J^{1,-1}_{(1)}$ can now easily be calculated. In order to
describe the final answer in a nice form introduce
\begin{eqnarray}
C & = & -\frac{4}{3}(J^{1,-1}_{(1)} + 3 (J^{0,0}_{(1)} )^2 ) \nonumber \\
E & = & J^{\frac{1}{2},-\frac{1}{2}}_{(1)} \nonumber \\
F & = & \frac{4}{3} \;
J^{\frac{1}{2},-\frac{1}{2}}_{(2)} \nonumber \\
H & = & 4 J^{0,0}_{(1)}  \label{gen}
\end{eqnarray}
(note that this is an invertible basis transformation). The Dirac
bracket algebra between these generators reads \cite{tjark}
\begin{eqnarray}
\{H,E\}^* & = & 2E \nonumber \\
\{H,F\}^* & = & -2F \nonumber \\
\{E,F\}^* & = & H^2+C   \label{rel1}
\end{eqnarray}
and $C$ Poisson commutes with everything. This algebra
which is called $\bar{W}_3^{(2)}$ was first
constructed in \cite{rocek} as a nonlinear deformation of
$su(2)$. In \cite{tjark} it was shown to be a reduction of
$sl_3$ and its representation theory was explicitly constructed.

Let's now consider the finite Miura transformation for this algebra.
Since the grading of $sl_3$  by $ad_{t_0}$ is half integer we have
to switch to the grading by $ad_{\delta}$. The explicit form of
$\delta$ is
\be
\delta = \frac{1}{3}\mbox{diag}(1,1,-2)
\ee
It is easily checked that this defines an integer grading of $sl_3$.
The crucial change is that the elements $E_{23}$ and $E_{12}$
have grade 1 and 0 w.r.t.
the $ad_{\delta}$ grading while they have grade $\frac{1}{2}$ w.r.t.
$ad_{t_0}$.
According to the prescription given in the previous section the
alternative set of constraints that one now imposes in order to
reduce the mixed system of first and second class constraints to
a system of first class constraints only is
\be
J^{1,1}_{(1)}-1=J^{\frac{1}{2},\frac{1}{2}}_{(2)}=0 \label{con}
\ee
As we already mentioned in the previous section, what has happened
here is that one has now imposed only half of the constraints that turned
out to be second class. The result is that both constraints in
(\ref{con}) are first class. The point however is that the
gauge symmetry induced by the second constraint in eq. (\ref{con})
can be completely fixed by adding the gauge fixing condition
\be
J^{\frac{1}{2},\frac{1}{2}}_{(1)}=0
\ee
which then leaves us with exactly the same set of constraints
and gauge invariances as before.

We can now describe the generalized Miura map for this case.
An arbitrary element of $t_++g_0$ is given by
\be
J_0 \equiv \left(
\begin{array}{ccc}
h+s  &  e  &  1  \\
f  &  s-h  & 0  \\
0  &  0  &  -2s
\end{array}
\right)
\ee
Note that $g_0 \simeq sl_2 \oplus u(1)$ which means that the Poisson
relations in $K(g_0)$ between the generators $h,e,f$ and $s$
(viewed as elements of $C^{\infty}(g_0)$) are given by
\begin{eqnarray}
\{h,e\} & = & e \nonumber \\
\{h,f\} & = & -f \nonumber \\
\{e,f\} & = & 2h   \label{rel2}
\end{eqnarray}
and $s$ commutes with everything.
As shown in the previous section the equation that we have to
solve in order to get explicit formulas for the Miura map is
is the following equation for $h \in \hat{H}$ (where $\hat{H}$ is
again the group of gauge transformations generated by the two
first class constraints (\ref{con})).
\be  \label{m}
h J_0 h^{-1}= \left(
\begin{array}{ccc}
J^{0,0}_{(1)} & 0 & 1 \\
J^{\frac{1}{2},-\frac{1}{2}}_{(2)} & -2J^{0,0}_{(1)} & 0 \\
J^{1,-1}_{(1)} & J^{\frac{1}{2},-\frac{1}{2}}_{(1)} & J^{0,0}_{(1)}
\end{array}
\right)
\ee
The unique solution of this equation is given by
\be
h= \left(
\begin{array}{ccc}
1 & 0 & 0 \\
0 & 1 & 0 \\
\frac{3}{2}s+\frac{1}{2}h & e & 1
\end{array}
\right)
\ee
Inserting this back into eqn.(\ref{m}) one finds certain expressions
for $J^{0,0}_{(1)},J^{\frac{1}{2},-\frac{1}{2}}_{(1)}
,J^{\frac{1}{2},-\frac{1}{2}}_{(2)}$ and $J^{1,-1}_{(1)}$ in
terms of the functions $\{e,f,h,s\}$. In terms of the generators
(\ref{gen}) these read
\begin{eqnarray}
H & = & 2h-2s \nonumber \\
E & = & (3s-h)e \nonumber \\
F & = & \frac{4}{3}\; f \nonumber \\
C & = & -\frac{4}{3}(h^2+3s^2+fe) \label{M}
\end{eqnarray}
It is easy to check, using the relations (\ref{rel2}) that these
satisfy the algebra (\ref{rel1}). Therefore we find that indeed
(\ref{M}) provides an injective Poisson homomorphism from
$W_3^{(2)}$ into the Kirillov Poisson algebra of the Lie algebra
$g_0=sl_2 \oplus u(1)$ \vspace{5mm}.

In the appendix we consider more examples
of finite $W$ algebras and their quantum versions.
In particular we construct all finite $W$ algebras associated to
$sl_4$ (to get the classical versions of the algebras displayed
in the appendix just replace commutators by Poisson brackets and
forget about all terms of order $\hbar^2$ or higher).

\newsection{Finite $W$ Symmetries in Generalized Toda Theories}

It is well known \cite{Gerv} that infinite $W$ algebras arise as
algebras of conserved currents of Toda theories. The infinite
$W$ algebras related to arbitrary $sl_2$ embeddings are related
to the conserved currents of so-called generalized Toda theories
\cite{LS,Dublin}. Since finite $W$ algebras are essentially a
dimensional reduction of infinite $W$ algebras, one expects that
there are one-dimensional analogues of the ordinary
two-dimensional Toda actions whose conserved currents are
related to
a finite $W$ algebra. Indeed, the construction of these
one-dimensional Toda actions is completely straightforward, and
has the desired properties.

One starts with the action for a free particle moving on the group
manifold $G=SL(n)$. The metric on $G$ is given by extending the
Cartan-Killing metric $(t_a,t_b)=\tr( t_at_b)$ on the Lie
algebra $g$ all over $G$ in a left-right invariant way.
This leads to the
familiar action
\be \label{freeac}
S[g]=\frac{1}{2} \int dt \tr \left(
g^{-1}\frac{dg}{dt} g^{-1}\frac{dg}{dt} \right).
\ee
Alternatively, this action can be seen as a one-dimensional WZW
action. It satisfies the following Polyakov-Wiegmann identity
\be \label{polwie}
S[gh]=S[g]+S[h]+\int dt \tr \left(
g^{-1} \frac{dg}{dt} \frac{dh}{dt} h^{-1} \right),
\ee
from which one immediately deduces the equations of motion,
\be \label{eqnmot}
\frac{d}{dt}
\left( g^{-1} \frac{dg}{dt} \right)=
\frac{d}{dt}
\left(  \frac{dg}{dt} g^{-1} \right)=0.
\ee
The action (\ref{freeac}) is invariant under $g\rightarrow h_1 g
h_2$ for constant elements $h_1,h_2\in G$. This leads to the
conserved currents $J,\bar{J}$ given by
\be \label{conscurr}
J
=\frac{dg}{dt}g^{-1} \equiv J^at_a
\,\,\,\, \mbox{\rm and} \,\,\,\,
\bar{J}
=g^{-1}\frac{dg}{dt} \equiv \bar{J}^at_a.
\ee
The equations that express the conservation of these currents in
time coincide with the equations of motion of the system, so
fixing the values of these conserved quantities completely fixes
the orbit of the particle once its position on $t=0$ is
specified. In this sense the free particle on a group is a
completely integrable system. The conserved quantities form  a
Poisson algebra \cite{papad}
\be \label{pois}
\{J^a,J^b\}=f^{ab}_{\,\,\, c}J^c,
\ee
with similar equations for $\bar{J}$. This is precisely the
Kirillov Poisson bracket we used as a starting point for the
construction of finite $W$ algebras. These were obtained by
imposing constraints on the Poisson algebra (\ref{pois}), and we
want to do the same here to get systems with finite $W$
symmetry. Actually, we already have the first example at our
disposal here. If we consider the trivial embedding of $sl_2$ in
$SL_n$, then the finite $W$ algebra is the Kirillov Poisson
algebra (\ref{pois}). The action (\ref{freeac}) is the
generalized Toda theory for the trivial embedding. The conserved
currents of this generalized Toda theory form a Poisson algebra
that is precisely the finite $W$ algebra associated to the
trivial embedding.

Finite $W$ algebras were obtained by imposing a set of first
class constraints. In terms of the decomposition $g=g_-\oplus
g_0 \oplus g_+$, these constraints were $\pi_{+}(J)=t_+$,
where $\pi_{\pm}$ are the projections on $g_{\pm}$. Here we want
to impose the same constraints, together with
similar constraints on
$\bar{J}$,
\be \label{constr}
\pi_+(J)=t_+\,\,\,\,\, \mbox{\rm and} \,\,\,\,\,
\pi_-(\bar{J})=t_-.
\ee
There are two equivalent ways to deal with these constraints.
One can either reduce the equations of motion, or reduce the
action for the free particle. For the two-dimensional case, both
were worked out in \cite{BFFOW,Dublin}. Let us first reduce the
equations of motion. If $G_{\pm}$ denote the subgroups of $G$
with Lie algebra $g_{\pm}$, and $G_0$ the subgroup with Lie
algebra $g_0$, then almost every element $g$ of $G$ can be
decomposed as $g_-g_0g_+$, where $g_{\pm,0}$ are elements of the
corresponding subgroups, because $G$ admits a generalized Gauss
decomposition $G=G_-G_0G_+$\footnote{Strictly speaking
$G_-G_0G_+$ is only dense in $G$ but we will ignore this subtlety
in the remainder}. Inserting $g=g_-g_0g_+$ into (\ref{constr})
we find
\ba \label{solco}
g_0^{-1} t_+ g_0 & = & \frac{dg_+}{dt} g_+^{-1} \nonu
g_0 t_- g_0^{-1} & = & g_-^{-1} \frac{dg_-}{dt}.
\ea
In the derivation of these equations one uses that
$\pi_+(g_-t_+g_-^{-1})=t_+$, and a similar equation with
$\pi_-$ and $t_-$, which follow from the fact that $t_{\pm}$
have degree $\pm 1$. The constrained currents look like
\ba \label{jconstr}
J & = & g_-\left(t_+ + \frac{dg_0}{dt} g_0^{-1} +
g_0 t_- g_0^{-1} \right) g_-^{-1}, \nonu
\bar{J} & = & g_+^{-1}\left(t_- + g_0^{-1} \frac{dg_0}{dt} +
 g_0^{-1} t_- g_0 \right) g_+.
\ea
The equations of motion now become
\ba \label{eqnm2}
0=g_-^{-1} \frac{dJ}{dt} g_- & = & \frac{d}{dt}
\left( \frac{dg_0}{dt} g_0^{-1} \right) + [g_0t_-g_0^{-1},t_+],
\nonu
0=g_+ \frac{d\bar{J}}{dt} g_+^{-1} & = & \frac{d}{dt}
\left( g_0^{-1} \frac{dg_0}{dt} \right) +
[t_-,g_0^{-1}t_+g_0].
\ea
which are generalized finite Toda equations as will be shown in a
moment.

Alternatively, one can reduce the action by writing
down the following one-dimensional version of a gauged
WZW model,
\begin{eqnarray} \label{wzwg}
S[g,A_+,A_-] & = & \frac{1}{2} \int dt \tr \left(
g^{-1}\frac{dg}{dt} g^{-1}\frac{dg}{dt} \right) \nonumber \\
&   & + \int dt \tr
\left( A_-(J-t_+)+A_+(\bar{J}-t_-)+A_-gA_+g^{-1} \right).
\end{eqnarray}
This action is invariant under the following transformations
\ba \label{invar}
g & \rightarrow & h_- g h_+,\nonu
A_- &  \rightarrow & h_-A_-h_-^{-1}-\frac{dh_-}{dt}h_-^{-1},
\nonu
A_+ &  \rightarrow & h_+^{-1}A_+h_+ -h_+^{-1}\frac{dh_+}{dt},
\ea
where $h_{\pm}$ are arbitrary elements of $G_{\pm}$.
We can use the gauge invariance to put $g_+=g_-=e$
(where $e$ is the unit element of the group $G$) in the Gauss
decomposition of $g$, thus we can take $g=g_0\in G_0$. Then from
the equations of motion for $A_{\pm}$ we find $A_+=g_0^{-1}t_+
g_0$ and $A_-=g_0t_- g_0^{-1}$. Substituting these back into the
action it reduces to
\be \label{finalact2}
S[g_0]  =  \frac{1}{2} \int dt \tr \left(
g_0^{-1}\frac{dg_0}{dt} g_0^{-1}\frac{dg_0}{dt} \right)
 - \int dt \tr
\left( g_0t_-g_0^{-1} t_+ \right).
\ee
The equations of motion for this action are indeed given by
(\ref{eqnm2}), showing the equivalence of the two approaches.

This generalized Toda action describes a particle moving on
$G_0$ in some background potential. Two commuting copies of the
finite $W$ algebra leave the action (\ref{finalact2}) invariant
and act on the space of solutions of the
equations of motion (\ref{eqnm2})\footnote{More precisely, the
symmetries of (\ref{finalact2}) form an algebra that is on-shell
isomorphic to a finite $W$ algebra}. This action is only given
infinitesimally, because we do not know how to exponentiate
finite (or infinite) $W$ algebras. One can, however, sometimes
find subspaces of the space of solutions that constitute a
minimal orbit of the $W$ algebra, see for example \cite{palla}
where this was worked out for the ordinary $W_3$ algebra.

For the principal embeddings of $sl_2$ in $sl_n$, the equations
of motion reduce to ordinary finite Toda equations of the type
\be \label{fintod}
\frac{d^2q_i}{dt^2}+\exp\left(\sum_{j=1}^{n-1}K_{ij}q_j\right)=0,
\ee
where $i=1,\ldots, n-1$, $K_{ij}$ is the Cartan matrix of
$sl_n$, and $g_0=\exp(q_iH_i)$.

The general solution of the the equations of motion
(\ref{eqnm2}) can be constructed as follows. Let
$h_0^{(1)},h_0^{(2)}$ be elements of $G_0$.
Let $X_0$ be an arbitrary element of $g_0$.
If $g_0(t)$ is defined by the Gauss decomposition
\be \label{allsol}
g_-(t)g_0(t)g_+(t) = h_0^{(1)} \exp t(X_0+(h_0^{(1)})^{-1}t_+
h_0^{(1)}+h_0^{(2)}t_-(h_0^{(2)})^{-1}) \,\,h_0^{(2)},
\ee
then $g_0(t)$ is the most general solution of (\ref{eqnm2}).
The easiest way
to find the action of the finite $W$ algebra on these solutions,
is to construct the conserved charges associated to these finite
$W$ symmetries (which can be done via a time dependent Miura
transformation), and to study the transformations they generate.
This might provide a valuable tool in the study of the solutions
(\ref{allsol}). We leave a detailed investigation of this, as
well as many other issues like
the quantization of the action (\ref{freeac}), to future study.

\newsection{Quantization of finite $W$ Algebras}

In quantum mechanics, quantization amounts to replacing Poisson
brackets by commutators. Since finite $W$ algebras are Poisson
algebras, the question arises whether it is possible to quantize
these Poisson algebras, to give finite quantum $W$ algebras. In
the infinite dimensional case (i.e. the usual infinite $W$
algebras), this is known to be possible
for the standard $W_n$ algebras associated to the principal embeddings.
The $W_3$ algebra
constructed by Zamolodchikov is a quantization of the Poisson
algebra one gets from hamiltonian reduction of
the affine $sl_3$ algebra. The most
difficult task in constructing infinite quantum $W$ algebras, is
to check that the resulting commutator satisfies the Jacobi
identity, or, equivalently, to check that the operator product
algebra is associative. Zamolodchikov did this explicitly for
his $W_3$ algebra. It is clear that this will become very
cumbersome for higher $W$ algebras, and that it is difficult to
obtain generic results in this direct approach.

A different way to find (infinite) quantum $W$ algebras has been
pioneered by Feigin and Frenkel \cite{FF}. In this approach
the quantum $W$ algebra is described as the zeroth cohomology of
a certain complex. The advantage of this approach is that one
automatically knows that the resulting operator algebra will be
associative. This procedure is closely related to BRST
quantization, and is usually called `quantum hamiltonian reduction'.
We will employ this method to study the quantization of finite $W$
algebras, related to arbitrary $sl_2$ embeddings of $sl_n$.
Another advantage of this method is that it provides a
functor from the category of representations of $g$ to those of
the quantum finite $W$ algebras, and is thus very useful to
study the representation theory of quantum finite $W$ algebras.

\newsubsection{Quantization}

Let $({\cal A}_0,\{.,.\} )$ be a commutative associative Poisson
algebra. A quantization of $({\cal A}_0,\{.,.\} )$ is an
associative algebra ${\cal A}$ depending on a parameter $\hbar$
such that (i) ${\cal A}$ is as a free $\ce[[\hbar]]$ module, (ii)
${\cal A}/\hbar {\cal A}\simeq {\cal A}_0$ and (iii) if $\pi$
denotes the natural map $\pi : {\cal A} \rightarrow {\cal
A}/\hbar {\cal A}\simeq {\cal A}_0$, then $\{\pi(X),\pi(Y)\}=
\pi((XY-YX)/\hbar)$. In most cases one has a set of generators
for ${\cal A}_0$, and ${\cal A}$ is completely fixed by giving
the commutation relations of these generators.

For example, let ${\cal A}_0$ be the Kirillov Poisson algebra of
polynomial
functions on a Lie algebra $g$, determined by equation (11).
Then a quantization of this Poisson algebra is the algebra
${\cal A}$
generated by the $J^a$ and $\hbar$, subject to the relations
$[J^a,J^b]=\hbar f^{ab}_{\,\,\, c} J^c$. Obviously, the Jacobi
identities are satisfied. Specializing to
$\hbar=1$, this algebra is precisely the universal enveloping
algebra ${\cal U}g$ of  $g$.

To find quantizations of finite $W$ algebras, one can first
reduce the $sl_n$ Kirillov Poisson algebra, and then try to
quantize the resulting algebras that we studied in the previous
sections. On the other hand, on can also first quantize and then
constrain. We will follow the latter approach, and thus study
the reductions of the quantum Kirillov algebra
\be \label{eq:qkir}
[J^a,J^b]=\hbar f^{ab}_{\,\,\, c} J^c.
\ee
We want to impose the same constraints on this algebra as we
imposed previously on the Kirillov Poisson algebra, to obtain
the quantum versions of the finite $W$ algebras related to
$sl_2$ embeddings. Imposing constraints on quantum algebras can
be done using the BRST formalism \cite{kost}. In the infinite
dimensional case, this has been done for the usual $W_N$
algebras by Feigin and Frenkel \cite{FF}. We use the
finite dimensional counterpart of this approach.

BRST quantization in the presence of second class
constraints is more cumbersome than in the presence of first
class constraints, it requires the introduction of extra
auxiliary fields to change the second class constraints into
first class constraints. However, it was shown above
that for arbitrary embeddings of $sl_2$ one can always choose a
set of constraints that is completely first class and leads to
the same $W$ algebras as the set $\{\phi^{j,m}_{(\mu)}\}$.
To perform a BRST quantization of the finite $W$ algebras we use
these alternative systems of first class constraints.

\newsubsection{The BRST Complex}

Consider the map $\chi : g_+\rightarrow \ce$ defined by
$\chi (E_{l+s_{k-1},l+s_k})=1$ for $l\geq1,1\leq k \leq n_l-1$
and $\chi (E_{ij})=0$ otherwise. Because the constraints
$\{J^{l+s_{k-1},r+s_k}-\delta^{r,l} \}_{l\geq 1;1\leq k \leq
n_l-1;r>0}$
are first class, $\chi$ defines a
one-dimensional representation of $g_+$.
In terms of $\chi$, the
constraints can be written as $\pi_+(J)=
\chi(\pi_+(J))$, where $\pi_+$ again denotes the
projection $g\rightarrow g_+$.
It is
this form of the constraints that we will use. Furthermore we
will take $\hbar=1$ for simplicity; the explicit $\hbar$
dependence can be determined afterwards.

As before latin indices will be
supposed to run over a basis ${t_a}$ of the Lie algebra $g$,
greek indices run over a basis of $g_+$ and
barred greek indices
(like $\bar{\alpha}$) run over a basis of
$g_-\oplus g_0$. Indices can again be raised and lowered by use of the
Cartan Killing metric.
The basis elements $t_a, t_{\alpha}$ and
$t_{\bar{\alpha}}$ are as before so chosen
that they have a well defined degree with respect to
$ad_{\delta}$.

To set up the BRST framework
we need to introduce anticommuting
ghosts and antighosts $c_{\alpha}$ and $b^{\alpha}$, associated
to the constraints that we want to impose \cite{kost}. They satisfy
$b^{\alpha}c_{\beta}+c_{\beta}b^{\alpha}=
\delta^{\alpha}_{\beta}$ and generate
the Clifford algebra $Cl(g_+\oplus g_+^*)$.
The quantum Kirillov algebra is
just the  universal enveloping algebra ${\cal U}g$,
and the total
space on which the BRST operator acts is $\Omega={\cal
U}g\otimes Cl(g_+\oplus g_+^*)$. A ${\bf Z}$
grading on $\Omega$ is defined by
$\deg(J^a)=0$, $\deg(c_{\alpha})=+1$ and $\deg(b^{\alpha})=-1$,
and we can decompose $\Omega=\oplus_k \Omega^k $ accordingly.
The BRST differential on $\Omega$ is given by $d(X)=[Q,X]$,
where $Q$ is the BRST charge
\be \label{def:q}
Q=(J^{\alpha}-\chi(J^{\alpha}))c_{\alpha}-\hf
f^{\alpha\beta}_{\,\,\, \gamma} b^{\gamma}c_{\alpha}c_{\beta}.
\ee
and $[.,.]$
denotes the graded commutator (as it always will from now on)
\be
[A,B]=AB- (-1)^{\mbox{deg}(A).\mbox{deg}(B)}BA
\ee
Note that $\mbox{deg}(Q)=1$.

This is the standard BRST complex associated to the first class
constraints of the previous section. Of interest are the
cohomology groups of this complex, $H^k(\Omega;d)$. The zeroth
cohomology group is the quantization of the classical finite $W$ algebra.
Because the gauge group $H$ in (39) acts
properly on $g_c$, we expect the higher cohomologies of the BRST
complex to vanish, as they are generically related to
singularities in the quotient $g_c/H$. In the mathematics
literature the cohomology of the BRST complex is called the Hecke
algebra ${\cal H}(g,g_+,\chi)$ associated to
$g,g_+,\chi$.
Hecke algebras related to arbitrary $sl_2$
embeddings have not been computed, apart from those
related to the principal $sl_2$ embeddings. In that case it was
shown by Kostant \cite{kost2} that the only nonvanishing
cohomology is $H^0(\Omega;d)$ and that it is isomorphic to the
center of the universal enveloping algebra. Recall that the
center of the ${\cal U}g$ is generated by the set of independent
Casimirs of $g$. This set is closely related to
the generators of standard infinite $W_n$-algebras; in that case
there is one $W$ field for each Casimir which
form a highly nontrivial algebra \cite{BBSS}. We see that for
finite $W$ algebras
the same generators survive, but that they form a trivial abelian
algebra. For non principal $sl_2$ embeddings however quantum finite $W$
algebras are non-trivial.

To compute the cohomology of $(\Omega;d)$, Feigin and Frenkel
make the crucial observation that the operator $d$ can be
decomposed into two commuting pieces. Write $Q=Q_0+Q_1$, with
\ba \label{def:q01}
Q_0 & = & J^{\alpha}c_{\alpha}-\hf
f^{\alpha\beta}_{\,\,\, \gamma} b^{\gamma}c_{\alpha}c_{\beta},
\nonu
Q_1 & = & -\chi(J^{\alpha})c_{\alpha},
\ea
and define $d_0(X)=[Q_0,X]$, $d_1(X)=[Q_1,X]$, then one can
verify by explicit computation that
$d_0^2=d_0d_1=d_1d_0=d_1^2=0$. Associated to this decomposition
is a bigrading of $\Omega=\oplus_{k,l} \Omega^{k,l}$ defined by
\ba \label{def:bigrade}
\deg(J^a)=(k,-k), & & \mbox{ if }t_a\in g_{k} \nonu
\deg(c_{\alpha})=(1-k,k), & & \mbox{ if }t_{\alpha} \in g_{k} \nonu
\deg(b^{\alpha})=(k-1,-k), & & \mbox{ if }t_{\alpha} \in g_{k},
\ea
with respect to which $d_0$ has degree $(1,0)$ and $d_1$ has
degree $(0,1)$. Thus $(\Omega^{k,l};d_0;d_1)$ has the structure
of a double complex. Explicitly, the action of $d_0$ and $d_1$
is given by
\ba \label{eq:q01}
d_0(J^a) & = & f^{\alpha a}_{\,\,\,b}J^bc_{\alpha},
\nonu
d_0(c_{\alpha}) & = & -\hf f^{\beta\gamma}_{\,\,\, \alpha}c_{\beta}
c_{\gamma}, \nonu
d_0(b^{\alpha}) & = & J^{\alpha} + f^{\alpha \beta}_{\,\,\,
\gamma}b^{\gamma}c_{\beta}, \nonu
d_1(J^a)=d_1(c_{\alpha}) & = & 0, \nonu
d_1(b^{\alpha}) & = & -\chi(J^{\alpha}).
\ea
To simplify the algebra, it is advantageous to introduce
\be \label{def:hj}
\hj^a=J^a+f^{a\beta}_{\,\,\, \gamma}b^{\gamma} c_{\beta}.
\ee
Our motivation to introduce these new elements $\hj^a$ is
twofold: first, similar expressions were encountered
in a study of the effective action for $W_3$ gravity \cite{BG},
where it turned out that the BRST cohomology for the infinite $W_3$
algebra case could conveniently be expressed in terms of
$\hj$'s; second, such expressions were introduced for the $J^a$'s
that live on the Cartan subalgebra of $g$ in \cite{FF}, and
simplified their analysis considerably. In terms of $\hj$ we
have
\ba \label{eq:q01hat}
d_0(\hj^a) & = & f^{\alpha a}_{\,\,\,\bar{\gamma}}
\hj^{\bar{\gamma}} c_{\alpha}, \nonu
d_0(c_{\alpha}) & = & -\hf f^{\beta\gamma}_{\,\,\, \alpha}c_{\beta}
c_{\gamma}, \nonu
d_0(b^{\alpha}) & = & \hj^{\alpha}, \nonu
d_1(\hj^a) & = & -f^{a \beta}_{\,\,\, \gamma} \chi(J^{\gamma})
c_{\beta}, \nonu
d_1(c_{\alpha}) & = & 0, \nonu
d_1(b^{\alpha}) & = & -\chi(J^{\alpha}).
\ea

The advantage of having a double complex, is that we can apply
the technique of spectral sequences \cite{botttu}
to it, in order to compute the cohomology of $(\Omega;d)$. The
results from the theory of spectral sequences that we need are
gathered in the next section.

\newsubsection{Spectral Sequences for Double Complexes}

Let $(\Omega^{p,q};d_0;d_1)$ denote a double complex, where $d_0$
has degree $(1,0)$ and $d_1$ has degree $(0,1)$. The standard
spectral sequence for this double complex is a sequence of
complexes $(E_r^{p,q};D_r)_{r\geq 0}$, where $D_r$ is a
differential of degree $(1-r,r)$, and is defined as follows:
$E_0^{p,q}=\Omega^{p,q}$, $D_0=d_0$, $D_1=d_1$, and for $r\geq 0$
\be \label{def:ss}
E^{p,q}_{r+1} = H^{(p,q)}(E_r;D_r)=
\frac{\ker\, D_r:E^{p,q}_r \rightarrow
E_r^{p+1-r,q+r}}{\im\, D_r:E^{p-1+r,q-r}_r\rightarrow E_r^{p,q}}.
\ee
The differential $D_{r+1}$ for $r>0$
is given by $D_{r+1}(\alpha)=d_1 (\beta)$,
where $\beta$ is chosen such that $d_0 \beta=D_r \alpha$. Such a
$\beta$ always exists and $D_{r+1}$ is uniquely defined in this
way. The usefulness of this spectral sequence is provided by the
following \cite{specseq}
\bt
If $\cap_{s}\oplus_{p,q\geq s}\Omega^{p,q}=\{0\}$,
then $E_{\infty}^{p,q}=\cap_r
E_r^{p,q}$ exists, and
 \be
\label{isom}
E_{\infty}^{p,q}\simeq\frac{F^q H^{(p+q)}(\Omega;d)}{F^{q+1}
H^{(p+q)}(\Omega;d)},
\ee
where
\be
\label{filter}
F^q H^{(p+q)}(\Omega;d)=\frac{\ker\ d:\oplus_{s\geq q}
\Omega^{p+q-s,s} \rightarrow \oplus_{s\geq q}
\Omega^{p+q+1-s,s}}{\im\ d:\oplus_{s\geq q} \Omega^{p+q-1-s,s}
\rightarrow \oplus_{s\geq q} \Omega^{p+q-s,s}}
\ee
\et
This spectral sequence is especially useful if it
collapses at the $n^{th}$ term for some $n$, \ie\ $D_r=0$ for
$r\geq n$, because then $E_n\simeq E_{\infty}$ and one needs only
to compute the first $n$ terms of the spectral sequence.

If the double complex is also an algebra, i.e. there is a
multiplication operator $m: \Omega^{p,q} \otimes_{\ce} \Omega^{p',q'}
\rightarrow \Omega^{p+p',q+q'}$, and $d$ satisfies the Leibniz
rule with respect to this multiplication, then (\ref{isom}) is
also an equality of algebras. It is in general  nontrivial
to reconstruct the full algebra structure of
$H^{\ast}(\Omega;d)$ from $E^{p,q}_{\infty}$, due to the
quotient in the right hand side of (\ref{isom}).

Another useful tool in the computation of cohomology is the
following version of the K\"unneth theorem
\bl
\label{kunneth}
Let $A=\oplus_k A^k$
be a graded differential algebra over $\ce$ with a
differential $d$ of degree 1 that satisfies the Leibniz rule, and
assume that $A$ has two graded subalgebras $A_1=\oplus_k A_1^k$
and $A_2=\oplus_k A_2^k$ such
that $d(A_1)\subset A_1$ and $d(A_2)\subset A_2$, that
$A_2^k=\{0\}$ for $k$ sufficiently large, and that
$m:A_1\otimes_{\ce} A_2 \rightarrow A$ given by $m(a_1\otimes
a_2)=a_1 a_2$ is an isomorphism of vector spaces. Then
\be \label{kunneth2}
H^{\ast}(A;d)\simeq \{a_1a_2 | a_1\in H^{\ast}(A_1;d), a_2\in
H^{\ast}(A_2;d) \}.
\ee
\el
Proof: form the double complex $(\Omega^{p,q};d_0;d_1)$ with
$\Omega^{p,q}\simeq m(A_1^p \otimes A_2^q)$, $d_0(a_1
a_2)=d(a_1)a_2$, and $d_1(a_1
a_2)=(-1)^{\deg(a_1)} a_1 d(a_2)$. The spectral sequence
for this double complex collapses at the $E_2$ term, and one
finds $E^{p,q}_{\infty}=\{a_1a_2|a_1 \in H^p(A_1;d), a_2\in
H^q(A_2;d)\}$. The condition $A_2^k=0$ for $k$ sufficiently
large guarantees that $F^q H^{p+q}=0$ for $q$ sufficiently
large, and one can assemble $H^{\ast}(\Omega;d)$ from
$E^{p,q}_{\infty}$ using (\ref{isom}). This leads to
(\ref{kunneth2}) on the level of vector spaces. Because $a_1a_2$
is really a representative of an element of $H^{\ast}(A;d)$, it
follows that (\ref{kunneth2}) is also an isomorphism of
algebras.\\

By induction, one can easily prove that the theorem still holds
if instead of two subalgebras $n$ subalgebras $A_1,\ldots A_n$
are given, with $A\simeq A_1 \otimes \ldots \otimes A_n$. The
condition $A_2^k=\{0\}$ for sufficiently large $k$ is replaced by
$A_2^k=\ldots =A_n^k=\{0\}$ for sufficiently large $k$. Without
such a condition, it may not be so easy to reassemble $H^{\ast}$
from $E_{\infty}$. To illustrate some of the difficulties that
can arise, let us give an example where $H^{\ast}$
cannot be recovered directly from $E_{\infty}$. This example is
not related to the above theorem, but it represents a situation
we will encounter in the computation of the BRST cohomology.

Consider the algebra $\Omega=\ce[x,y]/(y^2=0)$, where $x$ is an even
generator of bidegree $(1,-1)$, and $y$ is an odd generator of
bidegree $(0,-1)$. The differentials $d_0,d_1$ are given by
$d_{0,1}(x)=0$, $d_0(y)=x$ and $d_1(y)=-1$. One immediately
computes $H^k(\Omega;d)=\ce[x]/(x-1)\ce[x]\simeq \ce$
for $k=0$, and $H^k=0$
otherwise. The spectral sequence associated to the double
complex collapses at the first term, and one finds
$E^{p,q}_{\infty}=\ce \delta_{p,0} \delta_{q,0}$. Because $F^1
H^{\ast}=0$, one deduces that $H^k(\Omega;d)\simeq \ce \delta_{k,0}$.
On the other hand, we could also have started with the mirror
double complex obtained by interchanging $d_0$ and $d_1$ and the
bigrading. Thus, we assign bidegree $(-1,1)$ to $x$ and bidegree
$(-1,0)$ to $y$. The spectral sequence associated to the mirror
double complex also collapses at the first term, but now one
finds $E^{p,q}_{\infty}=0$. This is not in conflict with the
previous computation, because we cannot a priori find a $q$ for
which $F^q H^{p+q}=0$, and we can only conclude that $F^q
H^{p+q}\simeq F^{q+1} H^{p+q}$. If we compute explicitly with
respect to this bigrading what $F^q H^{p+q}$ is, we find that
it is only nontrivial for $p+q=0$, and then $F^q
H^0=x^q\ce[x]/(x-1)x^q\ce[x]\simeq\ce$ for $q\geq0$, and $F^q
H^0=\ce[x]/(x-1)\ce[x]\simeq\ce$ for $q<0$. This indeed yields
$E^{p,q}_{\infty}=0$. The lesson is that one should be careful
in deriving $H^{\ast}(\Omega;d)$ from $E^{p,q}_{\infty}$.

Finally, let us present another fact that will be useful later.
\bl \label{filter2}
Suppose $A$ is a differential graded algebra, $A=\oplus_{n\geq
0} A^n$, with a differential of degree $1$. Assume furthermore
that $A$ has a filtration
\be \label{filt}
\{0\}=F^0 A \subset F^1 A \subset F^2 A \subset \cdots \subset A,
\ee
such that $F^pA F^qA\subset F^{p+q}A$, and that $d$ preserves the
filtration, $d(F^pA)\subset F^pA$. If $H^k(F^{p+1}A/F^pA;d)=0$
unless $k=0$, then we have the following isomorphism of vector
spaces
\be \label{isom2}
H^0(A;d)\simeq \oplus_{p\geq 0} H^0(F^{p+1}A/F^pA;d).
\ee
\el
Proof: One can assign a spectral sequence to such a filtered
graded algebra \cite{specseq}, whose first term contains the
cohomologies $H^k(F^{p+1}A/F^pA;d)$. If only $H^0\neq 0$, then
the spectral sequence collapses at the first term, and because
the filtration is bounded from below ($\{0\}=F^0A$), one can
collect the vector spaces that make up $E^{\ast,\ast}_{\infty}$,
to get the isomorphism (\ref{isom2}).

\newsubsection{The BRST Cohomology}

The computation of the BRST cohomology is simplified
considerably due to the introduction of the new set of generators
$\hj^a$. The simplification arises due to
\bt
\label{reduc}
If $(\Omega;d)$ denotes the BRST complex, with $\Omega$
generated by $\hj^a$, $c_{\alpha}$ and $b^{\alpha}$, and $d=d_0+d_1$
given by (\ref{eq:q01hat}), then $H^{\ast}(\Omega;d)\simeq
H^{\ast}(\Omega_{red};d)$, where $\Omega_{red}$ is
the subalgebra of $\Omega$ generated by
$\hj^{\bar{\alpha}}$ and $c_{\alpha}$.
\et
Proof: Apply the K\"unneth theorem \ref{kunneth} to
$\Omega_{red}\otimes (\otimes_{\alpha} \Omega_{\alpha})$, where
$\Omega_{\alpha}$ is the algebra generated by $\hj^{\alpha}$ and
$b^{\alpha}$. Note that $[\hj^{\alpha},b^{\alpha}]=0$ and that
the conditions of the K\"unneth theorem are satisfied. Therefore,
$H^{\ast}(\Omega;d)\simeq H^{\ast}(\Omega_{red};d)\otimes
(\otimes_{\alpha} H^{\ast}(\Omega_{\alpha};d))$. Now
$(\Omega_{\alpha};d)$ is essentially the same complex as the
one we examined in the last part of the previous section, and one
easily proves that $H^k(\Omega_{\alpha};d)\simeq
\ce\delta_{k,0}$. This shows $H^{\ast}(\Omega;d)\simeq
H^{\ast}(\Omega_{red};d)$\\

The reduced complex $(\Omega_{red};d)$ is described by the
following set of relations:
\ba \label{eq:q01hat2}
d_0(\hj^{\bar{\alpha}}) & = & f^{\alpha \bar{\alpha}}_{\,\,\,
\bar{\gamma}} \hj^{\bar{\gamma}} c_{\alpha}, \nonu
d_0(c_{\alpha}) & = & -\hf f^{\beta\gamma}_{\,\,\, \alpha}c_{\beta}
c_{\gamma}, \nonu
d_1(\hj^{\bar{\alpha}}) & = & -f^{{\bar{\alpha}}
\beta}_{\,\,\, \gamma} \chi(J^{\gamma})
c_{\beta}, \nonu
d_1(c_{\alpha}) & = & 0, \nonu
[\hj^{\bar{\alpha}},\hj^{\bar{\beta}}]
& = & f^{\bar{\alpha}\bar{\beta}}_{
\,\,\,\bar{\gamma}} \hj^{\bar{\gamma}}, \nonu
[\hj^{\bar{\alpha}},c_{\beta}] & = &
-f^{\bar{\alpha}\gamma}_{\,\,\,\beta} c_{\gamma}, \nonu
[c_{\alpha},c_{\beta}] & = & 0.
\ea

Feigin and Frenkel \cite{FF} propose to use the spectral
sequence for the double complex $(\Omega;d_0;d_1)$ to compute
the BRST cohomology for the infinite case, to obtain the
standard $W$ algebras. They claim that in that case, the
spectral sequence collapses at the second term, and use this
to identify the $W$ algebras as the centralizers of some vertex
operators in a free field algebra. Because finite $W$ algebras
are the same as infinite $W$ algebras, with all dependence on the
co-ordinates suppressed (so that derivatives vanish), we would
expect the same thing to happen for finite $W$ algebras.
However, this turns out not to be the case here.

Let us demonstrate what happens for the case of the principal
$sl_2$ embedding in $sl_2$, \ie\ the embedding is given by
the identity map. The reduced algebra $\Omega_{red}$ is
generated by $\hat{H}$ of degree $(0,0)$, $\hat{F}$ of degree
$(-1,1)$ and $c$ of degree $(0,1)$. The nontrivial relations
between these generators are $[\hat{H},\hat{F}]=-2 F$ and
$[\hat{H},c]=-2c$. Furthermore, $d_1(\hat{F})=d_1(c)=0$,
$d_1(\hat{H})=-2c$, $d_0(\hat{H})=d_0(c)=0$ and
$d_0(\hat{F})=\hat{H}c$. To find $E^{\ast,\ast}_1$ we compute
$d_0(\hat{H}^a\hat{F}^bc)=0$ and
\be \label{se1}
d_0(\hat{H}^a\hat{F}^b)=b\hat{H}^a(\hat{H}+b-1)\hat{F}^{b-1}c.
\ee
Therefore in $H^{\ast}_{d_0}$ the identity
$f(\hat{H})\hat{F}^bc=f(1-b)\hat{F}^bc$ is valid, and
\be \label{sse1}
E^{\ast,\ast}_1=\ce[\hat{H}]\oplus\ce[\hat{K}]c.
\ee
Here we already disagree with \cite{FF}, because they claim
$E_1$ is only generated by $\hat{H}$ and $c$.
Next we compute the $E_2$ term of the spectral sequence. In
$E_1$ we have
\ba \label{se2}
d_1(f(\hat{H})) & = & (f(\hat{H})-f(\hat{H}+2))c \nonu
		& = & (f(0)-f(2))c,
\ea
from which it follows that
\be \label{sse2}
E_1^{\ast,\ast} = \{f(\hat{H})|f(0)=f(2)\}\oplus
\hat{F}\ce[\hat{F}]c.
\ee
This does not yet look like the final answer \cite{kost2}, namely
$H^k=0$ for $k\neq 0$ and $H^0$ is isomorphic to the center of
${\cal U}g$, generated by the second casimir of
$sl_2$. So let us compute $D_2$ to see whether the
spectral sequence has already collapsed. Using the definition of
$D_2$ given at the beginning of the previous section, we compute
subsequently
\ba \label{se3}
d_1(f(\hat{H}))
& = & (f(\hat{H})-f(\hat{H}+2))c, \nonu
d_0\left(\frac{f(\hat{H})-f(\hat{H}+2)}{\hat{H}}\hat{F}\right)
& = & (f(\hat{H})-f(\hat{H}+2))c, \nonu
d_1\left(\frac{f(\hat{H})-f(\hat{H}+2)}{\hat{H}}\hat{F}\right)
& = & \left(
\frac{f(\hat{H})-f(\hat{H}+2)}{\hat{H}} -
\frac{f(\hat{H}+2)-f(\hat{H}+4)}{\hat{H}+2} \right) \hat{F} c
\nonu
& = & (f(3)-f(-1))\hat{K}c.
\ea
and see that $D_2(f(\hat{H}))=(f(3)-f(-1))\hat{K}c$. The
spectral sequence does not collapse, and the next term in the
sequence is
\be \label{sse3}
E_3^{\ast,\ast}=\{f(\hat{H})|f(0)=f(2)\wedge f(-1)=f(3)\}\oplus
\hat{F}^2\ce[\hat{F}]c.
\ee
Continuing in this way one finds for the next terms in the
spectral sequence
\be \label{sen}
E_r^{\ast,\ast}=\{f(\hat{H})|f(-l)=f(l+2), 0\leq l \leq r-2\}
\oplus \hat{F}^{r-1}\ce[\hat{F}]c,
\ee
and finally
\be \label{seinf}
E_{\infty}^{\ast,\ast}=\{f(\hat{H})|f(1+\hat{H})=f(1-\hat{H})\}=
\ce[(\hat{H}-1)^2].
\ee
This agrees with the result of Kostant \cite{kost2}. The
spectral sequence does not collapse at all, and it is clear that
this is a rather cumbersome procedure to compute the BRST
cohomology.
Luckily, there is another spectral sequence one can associate to
a double complex, namely the sequence associated to
the mirror double complex obtained by interchanging the
bigrading and $d_0$ and $d_1$. This
spectral sequence turns out to be much
simpler, and will be examined in the next section, where we use
it to compute the BRST cohomology for arbitrary $sl_2$
embeddings.

\newsubsection{The Mirror Spectral Sequence}

The main result of this section is
\bt  \label{final}
As before let $g_{lw}\subset g$ be the kernel of the map
$\ad{t_-}:g\rightarrow g$. Then the BRST cohomology is
given by the following isomorphisms of vector spaces
\be \label{coho}
H^{k}(\Omega;d)\simeq ({\cal U}g_{lw})\delta_{k,0}.
\ee
\et
Proof: The $E_1$ term of the mirror spectral sequence is given
by the $d_1$ cohomology of $\Omega_{red}$. To compute the
cohomology we use lemma \ref{filter2}. The filtration on
$\Omega_{red}$ is: $F^p\Omega_{red}$ is spanned as a vector
space by $\{ \hj^{\bar{\alpha}_1} \hj^{\bar{\alpha}_2} \cdots
\hj^{\bar{\alpha}_r} c_{\beta_1}
 c_{\beta_2} \cdots c_{\beta_s} | r+s\leq p\}$. Thus
 $F^p\Omega_{red}/F^{p-1}\Omega_{red}$ is spanned by the
 products of precisely $p$ $\hj$'s and $c$'s, and in this
 quotient $\hj$ and $c$ (anti)commute with each other. Now let
 us rewrite $d_1(\hj^{\bar{\alpha}})$ as
 \ba
 \label{rewri}
 d_1(\hj^{\bar{\alpha}}) & = &
 -\tr([\chi(J^{\gamma})t_{\gamma},t^{\bar{\alpha}}] t^{\beta}
 c_{\beta} ) \nonu
 & =  & -\tr([t_+,t^{\bar{\alpha}}]t^{\beta}c_{\beta} ).
 \ea
 From this it is clear that $d_1(\hj^{\bar{\alpha}})=0$ for
 $t^{\bar{\alpha}}\in g_{hw}$. Furthermore, since $t_{\bar{\alpha}}\in
 g_0\oplus g_-$ and $\dim(g_{lw})=\dim(g_0)$, it
 follows that for each $\beta$ there is a linear combination
$a(\beta)_{\bar{\alpha}}\hj^{\bar{\alpha}}$ with
 $d_1(a(\beta)_{\bar{\alpha }}\hj^{\bar{\alpha}})=c_{\beta}$.
 This proves that
 \be \label{symalg}
 \bigoplus_{p>0}\frac{F^p\Omega_{red}}{F^{p-1}\Omega_{red}} \simeq
 \bigotimes_{t_{\bar{\alpha}}\in g_{lw}} \ce[\hj^{\bar{\alpha}}]
 \bigotimes_{t_{\bar{\alpha}}\not\in g_{lw}}
 (\ce[\hj^{\bar{\alpha}}] \oplus d_1(\hj^{\bar{\alpha}})
 \ce[\hj^{\bar{\alpha}}]).
 \ee
Using the K\"unneth theorem (lemma \ref{kunneth}) for
(\ref{symalg}), we find that
 \be \label{E1}
 H^k(\Omega_{red};d_1) =
 \bigotimes_{t_{\bar{\alpha}}\in g_{lw}}
\ce[\hj^{\bar{\alpha}}]\delta_{k,0} =
 ({\cal U}g_{lw})\delta_{k,0}.
 \ee
 Because there is only cohomology of degree 0, the mirror spectral
 sequence collapses, and $E_{\infty}= E_1$. Because
 $\Omega_{red}^{k,l}=0$ for $l>0$, we can find
 $H^{\ast}(\Omega_{red};d)$ from $E_{\infty}$. The theorem now follows
 directly from theorem \ref{reduc}.

\newsubsection{Reconstructing the Quantum Finite $W$ Algebra}

We succeeded in computing the BRST cohomology on the level of
vector spaces; as expected, there is only cohomology of degree
zero, and furthermore, the elements of $g_{lw}$ are in one-to-one
correspondence with the components of $g$ that made up the
highest weight gauge in section 1. Therefore
$H^{\ast}(\Omega;d)$ really is a quantization of the finite $W$
algebra. What remains to be done is to compute the algebra
structure of $H^{\ast}(\Omega;d)$. The only thing that
(\ref{coho}) tells us is that the product of two elements $a$
and $b$ of bidegree $(-p,p)$ and $(-q,q)$ is given by the
product structure on ${\cal U}g_{lw}$, modulo terms of bidegree
$(-r,r)$ with $r<p+q$. To find these lower terms we need
explicit representatives of the generators of
$H^0(\Omega;d)$ in $\Omega$. Such representatives can be
constructed using the so-called tic-tac-toe construction
\cite{botttu}: take some $\phi_0\in g_{lw}$, of bidegree $(-p,p)$.
Then $d_0(\phi)$ is of bidegree $(1-p,p)$. Because
$d_1d_0(\phi)=-d_0d_1(\phi)=0$, and there is no $d_1$ cohomology of
bidegree $(1-p,p)$, $d_0(\phi)=d_1(\phi_1)$ for some $\phi_1$ of
bidegree $(1-p,p-1)$. Now repeat the same steps for $\phi_1$,
giving a $\phi_2$ of bidegree $(2-p,p-2)$, such that
$d_0(\phi_1)=d_1(\phi_2)$. Note that $d_1d_0(\phi_1)=
-d_0d_1(\phi_1)=-d_0^2(\phi)=0$. In this way we find a sequence
of elements $\phi_l$ of bidegree $(l-p,p-l)$. The process stops
at $l=p$. Let $W(\phi)=\sum_{l=0}^{p} (-1)^l \phi_l$. Then
$dW(\phi)=0$, and $W(\phi)$ is a representative of $\phi_0$ in
$H^0(\Omega;d)$. The algebra structure of $H^0(\Omega;d)$ is
then completely determined by looking at the commutation
relations of $W(\phi)$ in $\Omega$, where $\phi_0$ runs over a
basis of $g_{lw}$. This is the quantum finite $W$ algebra.

Let us now give an example of the construction described above.

\newsubsection{Example}

Consider again the embedding associated to the following partition
of the number 3: $3=2+1$. We constructed the classical $W$ associated
to this embedding earlier. We shall now quantize this Poisson algebra
by the methods developed above.
Take the following basis of $sl_3$:
\be \label{basis21}
r_at_a=\mats{\frac{r_4}{6}-\frac{r_5}{2}}{r_2}{r_1}{r_6}{-
       \frac{r_4}{3}}{r_3}{r_8}{r_7}{\frac{r_4}{6}+\frac{r_5}{2}}.
\ee
Remember that
(in the present notation)
the $sl_2$ embedding is given by $t_+=t_1$, $t_0=-t_5$ and
$t_-=t_8$. The nilpotent subalgebra $g_+$ is spanned by
$\{t_1,t_3\}$, $g_0$ by $\{t_2,t_4,t_5,t_6\}$ and $g_-$ by
$\{t_7,t_8\}$. The $d_1$ cohomology of $\Omega_{red}$ is
generated by $\{\hj^4,\hj^5,\hj^6,\hj^8\}$, and using the
tic-tac-toe construction one finds representatives for these
generators in $H^0(\Omega_{red};d)$:
\ba \label{reps21}
W(\hj^4) & = & \hj^4, \nonu
W(\hj^6) & = & \hj^6, \nonu
W(\hj^7) & = & \hj^7-\hf\hj^2\hj^5-\hf\hj^4\hj^2+\hf\hj^2, \nonu
W(\hj^8) & = & \hj^8+\deel{1}{4}\hj^5\hj^5+\hj^2\hj^6-\hj^5.
\ea
Let us introduce another set of generators
\ba \label{newgens21}
C & = & -\deel{4}{3} W(\hj^8)-\deel{1}{9}W(\hj^4)W(\hj^4)-1,
\nonu
H & = & -\deel{2}{3}W(\hj^4)-1, \nonu
E & = &  W(\hj^7), \nonu
F & = & \deel{4}{3} W(\hj^6).
\ea
The commutation relations between these generators are given by
\ba \label{alg21}
[H,E] & = & 2E, \nonu
[H,F] & = & -2F, \nonu
[E,F] & = & H^2+C, \nonu
[C,E]=[C,F]=[C,H] & = & 0.
\ea
These are precisely the same as the relations for the
finite $W_3^{(2)}$ algebra given in \cite{tjark}.
Notice that in this case the quantum relations are identical to the
classical ones.
The explicit
$\hbar$ dependence can be recovered simply by multiplying the
right hand sides of (\ref{alg21}) by $\hbar$.

In the appendix we also discuss the explicit quantization of all
the finite $W$ algebras that can be obtained from $sl_4$.
There one does encounter certain quantum effects, i.e the quantum
relations will contain terms of order $\hbar^2$ or higher.

\newsection{The Representation Theory of Finite $W$ algebras}

The next important topic in the theory of finite $W$ algebras
is their representation theory. This representation theory will
presumably play an important role in the representation theory of
ordinary $W$ algebras as was already mentioned in the introduction.
It will be possible to construct the finite $W$ representations
by a quantum version of the generalized Miura map. This will give us
the ${\cal W}(i)$ representations as the Miura transform of $g_0$
representations.

If we denote by $X^{0,0}$ the component of an element $X$ of
bidegree $(0,0)$, so that $W^{0,0}(\phi)=(-1)^p\phi_p$, then
$[W^{0,0}(\phi),W^{0,0}(\phi')]=[W(\phi),W(\phi')]^{0,0}$, and
therefore $W(\phi) \rightarrow W(\phi)^{0,0}$ is a homomorphism of
algebras. We now have the following important theorem which is a
quantum version of the generalized Miura map.
\bt \label{qmiura} (Quantum Miura Transformation)
The map $W(\phi) \rightarrow W(\phi)^{0,0}$, or, equivalently,
the map $H^0(\Omega;d) \rightarrow H^0(\Omega;d)^{0,0}$, is an
isomorphism of algebras.
\et
Proof: Now that we know the cohomology of $(\Omega;d)$, let us
go back to the original double complex $(\Omega_{red};d_0;d_1)$.
Because there is only cohomology of degree $0$, we know that the
$E^{p,q}_{\infty}$ term of the spectral sequence associated to
$(\Omega_{red};d_0;d_1)$ must vanish unless $p+q=0$. If we look
at $d_0(\hj^{\bar{\alpha}})=f^{\beta\bar{\alpha}}_{\,\,\,
\bar{\gamma}}\hj^{\bar{\gamma}}c_{\beta}$, we see that
$d_0(\hj^{\bar{\alpha}})=0$ if and only if $\bar{\alpha}\in
g_0^{\ast}$. From this it is not difficult, repeating arguments
similar as those in the proof of theorem \ref{final},
to prove that the only nonvanishing piece of $\oplus_r
E^{r,-r}_1$ is in $E^{0,0}_1$. This implies that
$E^{p,q}_{\infty}$ is only nonzero for $p=q=0$. Because
$H^0(\Omega;d)=E^{0,0}_{\infty}=H^0(\Omega;d)^{0,0}$ as vector
spaces, it follows that the map $H^0(\Omega;d)\rightarrow
H^0(\Omega;d)^{0,0}$ can have no kernel, and is an isomorphism.

The quantum Miura transformation gives a faithful realization
of the quantum $W$ algebra in ${\cal U}g_0$.
As $g_0$ is nothing but a direct sum of simple Lie algebras (up to
$u(1)$ terms) its representation theory is just the standard
representation theory of (semi)simple Lie algebras. If $\rho$
is a representation of $g$ then the composition of $\rho$ with the
Miura map is a representation of the finite $W$ algebra. In this
way we get the representation theory of finite $W$ algebras from
the representation theory of the grade zero subalgebras associated
to the different $sl_2$ embeddings.

Since ${\cal U}(g_0)$
is abelian for the principal $sl_2$ embeddings, this
implies that in those cases the quantum finite $W$ algebras are
also abelian, something which was already proven by Kostant \cite{kost2}.
To get some interesting novel structure, one should therefore
consider nonprincipal $sl_2$
embeddings.

Again let us consider the example of $3=2+1$.
The expressions for the quantum Miura transformation of this
algebra
are obtained from (\ref{reps21}) by restricting these
expressions to the bidegree $(0,0)$ part. If we introduce
$s=(\hj^4+3\hj^5)/4$, $h=(\hj^5-\hj^4)/4$, $f=2\hj^6$ and
$e=\hj^2/2$, then $h,e,f$ form an $sl_2$ Lie
algebra, $[h,e]=e$ and $[h,f]=-f$ and $[e,f]=2h$ while $s$ commutes with
everything.
In terms of $s$ and $h,e,f$, the quantum Miura
transformation reads
\ba \label{miura21}
C & = & -\deel{4}{3}(h^2+\frac{1}{2}ef+\frac{1}{2}fe)-\deel{4}{9}s^2+
\deel{4}{3}s-1, \nonu
H & = & 2h-\deel{2}{3}s+1, \nonu
E & = & -2(s-h-1)e, \nonu
F & = & \deel{2}{3}f.
\ea
Notice that $C$ contains the second Casimir of $sl_2$, which is
what one would expect because it commutes with everything.
Every $g_0=sl_2\oplus u(1)$ module gives, using
the expressions (\ref{miura21}), a module for the finite quantum
algebra $W_3^{(2)}$.  So if we have a representation of $sl_2 \oplus u(1)$
in terms of $n \times n$ matrices, we immediately get a representation
of $\bar{W}^{(2)}_3$ in terms of $n \times n$ matrices.

In the appendix we give explicit formulas for the quantum Miura
transformations associated to the finite $W$ algebras that can be
obtained from $sl_4$.

\newsubsection{Fock realizations of quantum finite $W$ algebras}

Using the quantum Miura transformation we can turn any Fock realization
of $g_0$ into a Fock realization of the corresponding finite $W$
algebra. In this section we briefly discuss  how we can obtain
Fock realizations for simple Lie algebras.
We will then show in the
example of $W_3^{(2)}$ that its finite dimensional representations
of can be realized as submodules of certain Fock modules.

As usual define an oscillator algebra $O$ to be an
algebra generated by operators $\{a_i,a^{\dagger}_i\}_{i=1}^{n}$
with the following commutation relations
\be
[a_i,a_j]=[a_i^{\dagger},a_j^{\dagger}]=0 \;\; ; \;\;
[a_i,a_j^{\dagger}]=\delta_{ij}
\ee
In physics terminology this algebra describes a collection of $n$
uncoupled harmonic oscillators.
The Hilbert space
of such a system
is isomorphic to the space
${\bf C}[z_1, \ldots ,z_n]$ of polynomials in $n$ variables,
and (the quantum operators) $a_i$ and $a_j^{\dagger}$ can be
realized as
\be
a_i=\frac{\partial}{\partial z_i}\;\;\;\; \mbox{and} \;\;\;\;
a_i^{\dagger}=z_i
\ee
This is called the 'Fock realization' of the (quantum) harmonic
oscillator algebra.
Since any polynomial $P$ can be reduced to a multiple of 1 by successive
application of $\{a_i\}$ we can construct any polynomial $Q$ out
of $P$ by the action of the universal enveloping algebra of the
oscillator algebra. This means that the Fock realization is irreducible.
A Fock realization of a Lie algebra $g$ is an injective homomorphism from
$g$ to the universal enveloping algebra of the oscillator
algebra.

As we shall need this later
we now discuss how to obtain a Fock realization for an arbitrary
simple Lie algebra $g$
(see for details \cite{Kostant}).

Let $G$ be a simple Lie group with Lie algebra $g$. Furthermore
let $B_-$ be its lower Borel subgroup and $N_+$ the group
associated to the positive roots. Consider the coset space
\be
B_-\backslash G=\{B_-g \mid g\in G\}
\ee
The triple $(G,B_-\backslash G,B_-)$ is a principal bundle with
total space $G$, basespace $B_-\backslash G$ and structure group
$B_-$. Let now $\chi:B_- \rightarrow \bf C$ be a 1-dimensional
representation  (a character) of the Lie group $B_-$, then we can
construct the associated complex line bundle ${ \cal L}_{\chi}$ over
$B_-\backslash G$
\be
{ \cal L}_{\chi}=G\times {\bf C}/ \sim
\ee
where $(b.g,\chi (b^{-1})z) \sim (g,z)$ for all $b \in B_-$.
Denote the set of holomorphic global sections of this bundle by
$\Gamma_{\chi}$. The group $G$ acts from the right on ${\cal L}_{\chi}$
as follows
\be
[(g,z)]\cdot g' =[(gg',z)]
\ee
and this induces of course an action of $G$ (and therefore of $g$)
on $\Gamma_{\chi}$.

Now let $H$ and $b_-$ be the Cartan subalgebra and
negative Borel subalgebra of $g$ respectively. The derivative
of $\chi$ in the unit element
\be
\phi_{\chi}(b)=\frac{d}{dt}\chi (e^{tx})|_{t=0} \;\;\;\;\; x \in g
\ee
is a one dimensional representation of $b_-$. However
this implies that $0=[\phi_{\chi}(h),\phi_{\chi}(e_{-\alpha})]=
-\langle \alpha , h \rangle \phi_{\chi} (e_{\alpha})$ where
$h \in H$ and $e_{-\alpha}$ is a rootvector of $g$, which means
that $\phi_{\chi}(e_{-\alpha})=0$ for all $-\alpha \in \Delta_+$.
Therefore $\phi_{\chi}$ determines really an element $\Lambda_{\chi}$
of $H^*$. It is also clear that all elements of $H^*$ can be
obtained in this way.

It was already said above that $\Gamma_{\chi}$ is a representation
space of $g$. In fact the Borel-Weil
theorem states that for
$\Lambda_{\chi}$ an dominant weight of $g$ the space
of holomorphic sections $\Gamma_{\chi}$ of ${\cal L}_{\chi}$ is isomorphic
to the
finite dimensional irreducible representation of $g$ with highest weight
$\Lambda_{\chi}$. If $\Lambda_{\chi}$ is not a dominant weight
then the complex line bundle ${\cal L}_{\chi}$ has no non-trivial
holomorphic
sections.

The construction that leads to the Fock realization of the Lie
algebra $g$ is very similar to the Borel-Weil construction. First
one makes use of the fact that $B_-\backslash G$ possesses a
canonical decomposition, called the Bruhat decomposition
\be
B_-\backslash G= \bigcup_{w \in W} \; C_w
\ee
where $W$ is the Weyl group of $g$. The spaces $C_w$ are called
'Schubert cells' and are isomorphic to ${\bf C}^n$. Denote
from now on the Schubert cel corresponding to the unit element
of $W$ by $Y$. Since $Y \subset B_-\backslash G$
the group $G$ acts on $Y$ by right multiplication.
It can also be shown that $Y$ is diffeomorphic to $N_+$. This
diffeomorphism is defined by assigning to $[g] \in Y$ the unique
element $x \in N_+$ such that $[x]=[g]$ (i.e. such that
$x=bg$ for some $b \in B_-$). The right $G$-action on $Y$ induces a right
$G$-action on $N_+$ via this diffeomorphism. Let's from now
on  identify $Y$ and $N_+$.

The set $B_-N_+$ is a submanifold of $G$ and the triple $(B_-N_+,Y,B_-)$
is again a principle bundle (in fact it is nothing but the restriction
of the bundle $(G,B_-\backslash G,B_-)$ to the cell $Y\subset B_-\backslash
G$. Let again $\chi$ be a character of $B_-$ and let $\bar{{\cal L}}_{\chi}$
be the associated complex line bundle over $Y$. Denote the space of
sections of this line bundle by $\bar{\Gamma}_{\chi}$. It is a standard
result (and easy to prove) that there exists a 1-1 correspondence between
$\bar{\Gamma}_{\chi}$ and the set $\bar{R}_{\chi}$ of
holomorphic functions on $B_-N_+$ such that
\be
f(b.g)=\chi^{-1}(b)f(g)    \label{prop}
\ee
where $b\in B_-$ and $g \in N_+$. Since obviously the space
of holomorphic functions on $N_+$
and $\bar{R}_{\chi}$ are isomorphic (a holomorphic
function $f$ on $N_+$ uniquely determines a holomorphic function $\hat{f}$ on
$B_-N_+$ with the property  (\ref{prop}) if we set $\hat{f}(b.g)=
\chi^{-1}(b)f(g)$) we find that
\be
\bar{\Gamma}_{\chi}\simeq Hol(Y)
\ee
In \cite{Kostant} Kostant showed that there exists a representation
of the Lie algebra $g$ of $G$ in terms of first order smooth differential
operators on the manifold $Y$. Since the space $Y$
is isomorphic to ${\bf C}^{|\Delta_+|}$ where $|\Delta_+|$ is the
number of positive roots of $g$ we can choose a set
of global coordinates $z\equiv \{z_{\alpha}\}_{\alpha \in \Delta_+}$ on $Y$.
The representation
\be
\sigma_{\Lambda}:g \rightarrow \mbox{Diff}(Y)
\ee
of $g$ on the set of differential operators
$Diff(Y)$ on $Y$
is then given by
\be
\sigma_{\Lambda}(x)=\xi (x)+h_{\Lambda }(x) \;\;\;\; x\in g
\ee
where $\xi (x)$ is a Killing vector of the right $G$-action on $Y$
\be
\left( \xi (x)f \right) (z)=\frac{d}{dt}f(z.e^{tx})|_{t=0}
\;\;\;\; z \in Y
\ee
and
\be
\left( h_{\Lambda}(x)f \right)(z)=\langle \Lambda_* , Ad_z x \rangle
f(z) \label{hlab}
\ee
Here $\Lambda_*$ is the trivial extension of $\Lambda \in H^*$
to $g$
and $f$ is some holomorphic function of $Y$.
Note that in (\ref{hlab}) we made use of the fact we have
identified $Y$ and $N_+$ or else $Ad_z$ would not make sense.

The subspace ${\bf C}[\{z_{\alpha}\}]$ of $Hol(Y)$ is
obviously invariant  under  $\sigma_{\Lambda}(g)$ which means
that we have obtained Fock realizations of the Lie algebra $g$
(one for each $\Lambda \in H^*$).

For $sl_2$ the Fock realizations $\sigma_{\Lambda}$ have the form
\begin{eqnarray}
\sigma_{\Lambda}(e) & = & \frac{d}{dz} \nonumber \\
\sigma_{\Lambda}(f) & = & (\Lambda,\alpha )z-z^2\frac{d}{dz} \nonumber \\
\sigma_{\Lambda}(h) & = & \frac{1}{2}(\Lambda ,\alpha )-z\frac{d}{dz}
\label{F}
\end{eqnarray}
where $\alpha$ is the root of $sl_2$. The expressions for arbitrary
$sl_n$ can be found in \cite{BMP1} and will not be given here.

Using the results described above one is always able to immediately
write down a Fock realization of the algebra $g_0$ (since it is
essentially a direct sum of $sl_k$ algebras). Then using the quantum
Miura transformation one thus arrives at a Fock realization of the
finite $W$ algebra in question. Let us now explicitly do this in
the example of $\bar{W}_3^{(2)}$. Inserting the expressions
(\ref{F}) into the Miura map (\ref{miura21}) one finds
\begin{eqnarray}
\sigma_{\Lambda}(H) & = & (\Lambda,\alpha )-\frac{2}{3}s+1-2z\frac{d}{dz}
\nonumber \\
\sigma_{\Lambda}(E) & = & 2(1-s+\frac{1}{2}(\Lambda ,\alpha ))
\frac{d}{dz}-2z\frac{d^2}{dz^2} \nonumber \\
\sigma_{\Lambda}(F) & = & \frac{2}{3}(\Lambda,\alpha )z-\frac{2}{3}z^2
\frac{d}{dz} \nonumber \\
\sigma_{\Lambda}(C) & = & -\frac{1}{3}(\Lambda,\alpha )^2-
\frac{2}{3}(\Lambda,\alpha )-\frac{4}{9}s^2+\frac{4}{3}s-1 \label{FW}
\end{eqnarray}
(where we consider $s$ to be a number). This realization is equal to
the zero mode structure of the free field realization of the infinite
$W_3^{(2)}$ algebra constructed in \cite{B}. Note however that the
derivation is completely different since in \cite{B} (using standard
methods) the expressions in terms of free fields were obtained by
constructing the generators of the commutant of certain screening
charges. Constructing this commutant is in general however rather
cumbersome. The method we presented above is more direct and
works for arbitrary embeddings (and realizations).

The Fock realizations (\ref{FW}) contain for certain values of
$s$ and $(\Lambda, \alpha )$  the finite dimensional representations
of $\bar{W}_3^{(2)}$ which were constructed in \cite{tjark}.
Before we show this let us recall the results of \cite{tjark}.
\bt
Let $d$ be a positive integer and $x$ a real number.
\begin{enumerate}
\item For every pair $(p,x)$ the algebra $\bar{W}_3^{(2)}$ has a
unique highest weight representation $W(d;x)$ of dimension $d$
with highest weight  $j(d;x)=d+x-1$ and central value
$c(d;x)=\frac{1}{3}(1-d^2)-x^2$.
\item Let $k\in \{1, \ldots , d-1 \}$ then
$W(d;\frac{2}{3}k-\frac{1}{3}d)$ is reducible and its invariant
subspace is isomorphic as a representation to $W(d-k;-\frac{1}{3}(k+d))$.
\item The representation $W(d;x)$ is unitary iff $x>\frac{1}{3}d-
\frac{2}{3}$
\end{enumerate}
\et
Now one can easily check that for $d=(\Lambda ,\alpha )$ and
$s=3-\frac{2}{3}x$ the subspace
\be
V=\{P(z) \in {\bf C}[z] \mid (\frac{d}{dz})^dP(z)=0 \}
\ee
of ${\bf C}[z]$ is isomorphic as a representation to $W(d;x)$.

\newsection{Discussion}
In this paper we have studied finite $W$ algebras in great detail and
it turns out that they are very rich in their structure.
There are several issues that deserve further study. For example
it would be very interesting to calculate the orbit of finite
$W$ transformations on the solution space of the finite dimensional
generalized Toda systems we encountered in this paper. These systems
were already derived in \cite{VeBa} as static and spherically symmetric
solutions of the self dual Yang-Mills equations. In that paper some
special solutions of generalized Toda theories were constructed but
as far as we know the general solution space is not known. Since
finite $W$ algebras act on this solution space, transforming one
solution into another, it may be possible to generate the
entire solution space by the finite $W$ action (remember that
the symmetry group of the free particle on the group also acts
transitively on the space of solutions).

Closely related to this problem is the problem of finding the
symplectic orbits of finite $W$ algebras (cf. \cite{sorb}, where
a characterization of these was given for the infinite standard
$W_N$ algebras). Remember that the
Kirillov Poisson structure is not associated to a symplectic
form but that the Lie algebra splits up into union of
symplectic orbits. It is well known that these orbits are just
the coadjoint orbits of the group action on the Lie algebra.
It may be interesting to apply the procedure of geometric quantization
to the symplectic orbits of classical finite $W$ algebras and see
if one can reproduce the representations $W(d;x)$. Of course
this would be equivalent to finding a Borel-Weil like theorem
for finite $W$ algebras.

Another interesting problem is finding comultiplications for
finite $W$ algebras, in order to be able to define tensor
products. This is a difficult problem, since the natural
comultiplication on the universal enveloping algebra does not
induce one on the $W$ algebras.

Many of the techniques developed in this paper can equally well
be applied to ordinary $W$ algebras. We will come back to this in
a future publication.

\newsection{Appendix: Finite $W$ Algebras from $sl_4$}
In this appendix we give some explicit results on the quantum
finite $W$ algebras associated to $sl_4$.

\newsubsection{$4= 2+ 1 + 1$}

The basis of $sl_4$ we use to study this quantum algebra is
\be \label{basis211}
r_at_a=\left(\begin{array}{cccc}
\frac{1}{2}r_{10}-\frac{1}{8}r_8 -\frac{1}{8}r_9 & r_{11} &
r_{12} & r_{15} \\
r_5 & \frac{3}{8}r_8-\frac{1}{8}r_9 & r_7 & r_{14} \\
r_4 & r_6 & -\frac{1}{8}r_8+\frac{3}{8}r_9 & r_{13} \\
r_1 & r_2 & r_3 &
-\frac{1}{2}r_{10}-\frac{1}{8}r_8 -\frac{1}{8}r_9
\end{array} \right).
\ee
The $sl_2$ embedding is given by $t_+=t_{15}$, $t_0=t_{10}$ and
$t_-=\hf t_1$. The nilpotent algebra $g_+$ is spanned by
$\{t_{13},t_{14},t_{15}\}$, $g_0$ by $\{t_4,\ldots,t_{12}\}$ and
$g_-$ by $\{t_1,t_2,t_3\}$. The $d_1$ cohomology of
$\Omega_{red}$ is generated by $\hj^a$ for $a=1, \ldots ,9$.
Representatives that are exactly $d$-closed, are given by
$W(\hj^a)=\hj^a$ for $a=4,\ldots,9$, and by
\ba \label{reps211}
W(\hj^1) & = & \hj^1 + \hj^4\hj^{12}+\hj^5\hj^{11}+\deel{1}{4}
\hj^{10}\hj^{10}+\deel{\hbar}{2}\hj^{10}, \nonu
W(\hj^2) & = & \hj^2 + \hj^6\hj^{12} + \hf\hj^8\hj^{11} +
\hf\hj^{10}\hj^{11}+\hbar\hj^{11}, \nonu
W(\hj^3) & = & \hj^3 + \hj^7\hj^{11} + \hf\hj^9\hj^{12} +
\hf\hj^{12}\hj^{10}+\deel{\hbar}{2}\hj^{12}.
\ea
Introduce a new basis of fields as follows
\ba \label{newgens}
U & = & \deel{1}{4} (\whj{8}+\whj{9}), \nonu
H & = & \deel{1}{4} (\whj{8}-\whj{9}), \nonu
F & = & -\whj{7}, \nonu
E & = & -\whj{6}, \nonu
G_1^- & = & -\whj{3},  \nonu
G_1^+ & = & \whj{2}, \nonu
G_2^- & = & \whj{5}, \nonu
G_2^+ & = & \whj{4}, \nonu
C & = & \whj{1}+\hf EF+\hf FE + H^2+\hf U^2 + 2\hbar U.
\ea
If we compute the commutators of these expressions, we find that
$C$ commutes with everything, $\{E,F,H\}$ form a $sl_2$
subalgebra and $G_i^{\pm}$ are spin $\hf$ representations for
this $sl_2$ subalgebra. $U$ represents an extra $u(1)$ charge.
The nonvanishing commutators, with $\hbar$ dependence,  are
\ba \label{alg211}
[E,F] & = & 2\hbar H , \nonu
[H,E] & = & \hbar E , \nonu
[H,F] & = & -\hbar F , \nonu
[U,G_1^{\pm}] & = & \hbar G_1^{\pm}, \nonu
[U,G_2^{\pm}] & = & -\hbar G_2^{\pm}, \nonu
[H,G_i^{\pm}] & = & \pm\deel{\hbar}{2} G_i^{\pm}, \nonu
[E,G_i^-] & = & \hbar G_i^+, \nonu
[F,G_i^+] & = & \hbar G_i^-, \nonu
[G_1^+,G_2^+] & = & -2\hbar E(U+\hbar), \nonu
[G_1^-,G_2^-] & = & 2\hbar F(U+\hbar), \nonu
[G_1^+,G_2^-] & = & \hbar(-C+EF+FE+2H^2+\deel{3}{2}U^2+2HU)+\hbar^2
(2H+3U), \nonu
[G_1^-,G_2^+] & = & \hbar(C-EF-FE-2H^2-\deel{3}{2}U^2+2HU)+\hbar^2
(2H-3U).
\ea
Let us also present the quantum Miura transformation for this
algebra. In this case, $g_0=sl_3\oplus u(1)$. Standard
generators of $g_0$ can be easily identified. A generator of
the $u(1)$ is $s=\hf\hj^8+\hf\hj^9+2\hj^{10}$, and the $sl_3$
generators are $e_1=\hj^5$, $e_2=\hj^6$, $e_3=\hj^4$,
$f_1=\hj^{11}$, $f_2=\hj^7$, $f_3=\hj^{12}$,
$h_1=-\hf\hj^8+\hf\hj^{10}$ and $h_2=\hf\hj^8-\hf\hj^9$.
The convention is such that the commutation relations between
$\{e_i,f_i,h_i\}$ are the same as those of corresponding
matrices defined by
\be \label{conv211}
a_i e_i + b_i f_i + c_i
h_i=\hbar\mats{c_1}{a_1}{a_3}{b_1}{c_2-c_1}{a_2}{b_3}{b_2}{-c_2}.
\ee
The quantum Miura transformation reads
\ba \label{miura211}
U & = & \deel{1}{6}(s-2h_2-4h_1), \nonu
H & = & \hf h_2, \nonu
F & = & -f_2, \nonu
E & = & -e_2, \nonu
G_1^- & =  & -f_2f_1-\deel{1}{3}(s-2h_2-h_1+3\hbar)f_3, \nonu
G_1^+ & = & e_2f_3+\deel{1}{3}(s+h_2-h_1+3\hbar)f_1, \nonu
G_2^- & = & e_1, \nonu
G_2^+ & = & e_3, \nonu
C & = & (\deel{1}{24}s^2+\deel{\hbar}{2}s) +
\hf(e_1f_1+f_1e_1+e_2f_2+f_2e_2+e_3f_3+f_3e_3) \nonu
& & +
\deel{1}{3} (h_1^2+h_1h_2+h_2^2).
\ea
In $C$ we again recognize the second casimir of $sl_3$. It is a
general feature of finite $W$ algebras that they contain a
central element $C$, whose Miura transform contains the second
casimir of $g_0$. $C$ is the finite counterpart of the energy
momentum tensor that every infinite $W$ algebra possesses.

\newsubsection{$4= 2+2$}

A convenient basis to study this case is
\be \label{basis22}
r_at_a=\left(\begin{array}{cccc} \frac{r_6}{4}+\frac{r_{10}}{2}
& \frac{r_5}{2}+\frac{r_8}{2} & r_{12} & r_{14} \\
\frac{r_7}{2}+\frac{r_9}{2} & -\frac{r_6}{4}-\frac{r_{11}}{2} &
r_{13} & r_{15} \\ r_1 & r_2 & \frac{r_6}{4}-\frac{r_{10}}{2} &
\frac{r_5}{2}-\frac{r_8}{2} \\ r_3 & r_4 &
\frac{r_7}{2}-\frac{r_9}{2} & -\frac{r_6}{4}+\frac{r_{11}}{2}
\end{array} \right).
\ee
The $sl_2$ embedding is given by $t_+=t_{12}+t_{15}$,
$t_0=t_{10}-t_{11}$ and $t_-=\hf(t_1+t_4)$. The subalgebra $g_+$
is spanned by $\{t_{12},\ldots,t_{15}\}$, $g_0$ by
$\{t_5,\ldots,t_{11}\}$ and $g_-$ by $\{t_1,\ldots,t_4\}$. The
$d_1$ cohomology of $\Omega_{red}$ is generated by
$\hj^1,\ldots,\hj^7$. The $d$-closed representatives are
$W(\hj^a)=\hj^a$ for $a=5,6,7$, and
\ba \label{reps22}
W(\hj^1) & = & \hj^1-\hv\hj^5\hj^9+\hv\hj^7\hj^8+\hv\hj^8\hj^9+
 \hv\hj^{10}\hj^{10}+\deel{3\hbar}{4}\hj^{10}-\deel{\hbar}{4}
 \hj^{11}, \nonu
 W(\hj^2) & = &
 \hj^2+\hv\hj^5\hj^{10}+\hv\hj^5\hj^{11}-\hv\hj^6\hj^8+
 \hv\hj^8\hj^{10}-\hv\hj^{10}\hj^{11}+\deel{\hbar}{2}\hj^8,
 \nonu
 W(\hj^3) & = &
 \hj^3+\hv\hj^6\hj^9-\hv\hj^7\hj^{10}-\hv\hj^7\hj^{11}+
 \hv\hj^9\hj^{10}-\hv\hj^9\hj^{11}+\deel{\hbar}{2}\hj^9,
 \nonu
W(\hj^4) & = &
\hj^4+\hv\hj^5\hj^9-\hv\hj^7\hj^8+\hv\hj^8\hj^9+
\hv\hj^{11}\hj^{11}+\deel{\hbar}{4}\hj^{10}-
\deel{3\hbar}{4}\hj^{11}.
\ea
To display the properties of the algebra as clearly as possible,
we introduce a new basis of fields
\ba \label{newgens22}
H & = & \hf\whj{6}, \nonu
E & = & -\whj{7}, \nonu
F & = & -\whj{5}, \nonu
G^+ & = & \whj{3}, \nonu
G^0 & = & \whj{1}-\whj{4}, \nonu
G^- & = & -\whj{2}, \nonu
C & = & \whj{1}+\whj{4}+\deel{1}{8}\whj{6}\whj{6}+
 \hv\whj{5}\whj{7} \nonu
 & & +\hv\whj{7}\whj{5}+\deel{\hbar}{4}\whj{6}.
\ea
Here, $C$ is the by now familiar central element, $\{E,H,F\}$
form an $sl_2$ algebra and $\{G^+,G^0,G^-\}$ form a spin $1$
representation with respect to this $sl_2$ algebra. The
nonvanishing commutators are
\ba \label{alg22}
[E,F] & = & 2\hbar H, \nonu
[H,E] & = & \hbar E, \nonu
[H,F] & = & -\hbar F, \nonu
[E,G^0] & = & 2\hbar G^+, \nonu
[E,G^-] & = & \hbar G^0, \nonu
[F,G^+] & = & \hbar G^0, \nonu
[F,G^0] & = & 2\hbar G^-, \nonu
[H,G^+] & = & \hbar G^+, \nonu
[H,G^-] & = & -\hbar G^-, \nonu
[G^0,G^+] & = & \hbar(-CE+EH^2+\hf EEF+\hf EFE)-2\hbar^3 E,
\nonu
[G^0,G^-] & = & \hbar(-CF+FH^2+\hf FFE+\hf FEF)-2\hbar^3 F,
\nonu
[G^+,G^-] & = & \hbar(-CH+H^3+\hf HEF+\hf HFE)-2\hbar^3 H.
\ea
Since $g_0=sl_2\oplus sl_2 \oplus u(1)$, the quantum Miura
transformation expresses this algebra in term of generators
$\{e_1,h_1,f_1\}$, $\{e_2,h_2,f_2\}$, $s$ of $g_0$. The relation
between these generators and the $\hj^a$ are:
$s=\hj^{10}-\hj^{11}$, $h_1=\hf(\hj^6+\hj^{10}+\hj^{11})$,
$h_2=\hf(\hj^6-\hj^{10}-\hj^{11})$, $e_1=\hf(\hj^7+\hj^9)$,
$e_2=\hf(\hj^7-\hj^9)$, $f_1=\hf(\hj^5+\hj^8)$ and
$f_2=\hf(\hj^5-\hj^8)$. The commutation relations for these are
$[e_1,f_1]=\hbar h_1$, $[h_1,e_1]=2\hbar e_1$,
$[h_1,f_1]=-2\hbar f_1$, and similar for $\{e_2,h_2,f_2\}$.
For the quantum Miura transformation one
then finds
\ba \label{miura22}
H & = & \hf(h_1+h_2), \nonu
E & = & -e_1-e_2, \nonu
F & = & -f_1-f_2, \nonu
G^+ & = & \hf e_1h_2-\hf e_2h_1 +\hv s(e_1-e_2)+\hbar(e_1-e_2),
\nonu
G^0 & = & f_1e_2-f_2e_1+\hv s (h_1-h_2)+\hbar(h_1-h_2), \nonu
G^- & = & \hf f_1h_2 -\hf f_2 h_1 -\hv s (f_1-f_2)-\hbar
(f_1-f_2), \nonu
C & = & (\deel{1}{8}s^2+\hbar s) +
\hf(e_1f_1+f_1e_1+e_2f_2+f_2e_2) +
\hv(h_1^2 +h_2^2).
\ea
The infinite dimensional version of this algebra is one of the
`covariantly coupled' algebras that have been studied in
\cite{tjark2}. The finite algebra (\ref{alg22})
is almost a Lie algebra. If we assign particular values to $C$
and to the second casimir $C_2=(H^2+\hf EF+\hf FE)$ of the $sl_2$
subalgebra spanned by $\{E,H,F\}$, then (\ref{alg22}) reduces to
a Lie algebra. For a generic choice of the values of $C$ and
$C_2$ this Lie algebra is isomorphic to $sl_2 \oplus sl_2$. An
interesting question is, whether a similar phenomena occurs
for different covariantly coupled algebras.

\newsubsection{$4=3+1$}

The last nontrivial nonprincipal $sl_2$ embedding we consider is
$\underline{4}_4\simeq \underline{3}_2\oplus \underline{1}_2$. We choose yet
another basis
\be \label{basis31}
r_at_a = \left( \begin{array}{cccc}
\frac{r_{5}}{12}-\frac{r_{6}}{3} & r_8 & r_{11} & r_{12} \\
r_4 & -\frac{r_5}{4} & r_{13} & r_{14} \\
\frac{r_3}{2}-\frac{r_9}{2} & r_{10} &
\frac{r_5}{12}+\frac{r_6}{6}+\frac{r_7}{2} & r_{15} \\
r_1 & r_2 & \frac{r_3}{2}+\frac{r_9}{2} &
\frac{r_5}{12}+\frac{r_6}{6}-\frac{r_7}{2}
\end{array} \right),
\ee
in terms of which the $sl_2$ embedding is $t_+=t_{11}+t_{15}$,
$t^0=-3t_6+t_7$ and $t_-=2t_3$. The subalgebra $g_+$ is
generated by $\{t_{11},\ldots,t_{15}\}$, $g_-$ is generated by
$\{t_1,t_2,t_3,t_9,t_{10}\}$, and $g_0$ is generated by
$\{t_4,\ldots,t_8\}$. The $d_1$ cohomology is generated by
$\hj^1,\ldots,\hj^5$, and $d$-closed representatives are given
by $\whj{4}=\hj^4$, $\whj{5}=\hj^5$, and
\ba \label{reps31}
\whj{1} & = & \hj^1+\www{1}{6}{3}{6}-\www{1}{12}{6}{3}
+\www{1}{4}{7}{3}+\hj^4\hj^{10}+\www{1}{4}{6}{9}-\www{1}{4}{7}{9}-
\www{1}{3}{4}{5}\hj^8 \nonu
& &
-\www{1}{3}{4}{6}\hj^8-\www{1}{108}{6}{6}
\hj^6+\www{1}{12}{6}{7}\hj^7+\www{31\hbar}{12}{4}{8}-
\deel{3\hbar}{4}\hj^9+\www{5\hbar}{48}{6}{6}
\nonu
& &
+\www{\hbar}{6}{7}{6}-\www{3\hbar}{16}{7}{7}-
\deel{3\hbar^2}{8}\hj^7-\deel{7\hbar^2}{24}\hj^6,
\nonu
\whj{2} & = & \hj^2-\www{1}{2}{3}{8}-\www{1}{3}{5}{10}-
\www{1}{6}{6}{10}+\www{1}{2}{7}{10}-\www{1}{2}{8}{9}+
\deel{3\hbar}{2}\hj^{10}+
\www{1}{9}{5}{5}\hj^8 \nonu
& &
+\www{1}{9}{5}{6}\hj^8+
\www{1}{36}{6}{6}\hj^8-\www{1}{4}{7}{7}\hj^8-
\hbar\hj^5\hj^8-\www{\hbar}{2}{6}{8}-\www{\hbar}{2}{7}{8}+
2\hbar^2\hj^8, \nonu
\whj{3} & = & \hj^3 +\hj^4\hj^8+\www{1}{12}{6}{6}+
\www{1}{4}{7}{7}+\deel{\hbar}{2}\hj^7-\deel{\hbar}{2}\hj^6.
\ea
We introduce a new basis
\ba \label{newgens31}
U & = & \deel{1}{4}\whj{5}, \nonu
G^+ & = & \whj{4}, \nonu
G^- & = & \whj{2}, \nonu
S & = & \whj{1}, \nonu
C & = &
\whj{3}+\deel{1}{24}\whj{5}\whj{5}-\deel{\hbar}{2}\whj{5}.
\ea
In this case, the fields are not organized according to $sl_2$
representations, because the centralizer of this $sl_2$
embedding in $sl_4$ does not contain an $sl_2$. Again $C$ is a
central element, and the nonvanishing commutators are
\ba \label{alg31}
[U,G^+] & = & \hbar G^+ , \nonu
[U,G^-] & = & -\hbar G^- , \nonu
[S,G^+] & = & \hbar G^+ (-\deel{2}{3} C+\deel{20}{9}U^2-
\deel{43\hbar}{9}U+\deel{29\hbar^2}{27}), \nonu
[S,G^-] & = & \hbar  (\deel{2}{3} C-\deel{20}{9}U^2+
\deel{43\hbar}{9}U-\deel{29\hbar^2}{27})G^-, \nonu
[G^+,G^-] & = & \hbar S-\deel{4\hbar}{3}CU+\deel{3\hbar^2}{4}C
    +\deel{88\hbar}{27}U^3-\deel{17\hbar^2}{2}U^2+
    \deel{25\hbar^2}{6}U.
\ea
This is the first example where the brackets are no longer
quadratic, but contain third order terms.
For the sake of completeness, let us also give the quantum Miura
transformation for this algebra. We identify generators
$\{e,f,h\}$,$s_1$,$s_2$ of $sl_2\oplus u(1) \oplus u(1)=g_0$ via
$f=\hj^8$, $e=\hj^4$, $h=\deel{1}{3}(\hj^5-\hj^6)$,
$s_1=\deel{1}{3}(2\hj^6+\hj^5)$ and $s_2=\hj^7$. The only
nontrivial commutators between these five generators are
$[e,f]=\hbar h$, $[h,e]=2\hbar e$ and $[h,f]=-2\hbar f$.
The quantum
Miura transformation now reads
\ba \label{miura31}
U & = & \deel{1}{4}s_1+\deel{1}{2}h, \nonu
G^+ & = & e, \nonu
G^- & = & (\deel{1}{4}s_1^2+\deel{1}{2}s_1 h + \deel{1}{4}h^2
-\deel{1}{4}s_2^2-\deel{\hbar}{2}(3s_1+3h+s_2)+2\hbar^2)f,
\nonu
S & = & -\deel{1}{12} e(8s_1+4h-31\hbar)f-\deel{1}{108}
(s_1-h)^3+\deel{1}{12}(s_1-h)s_2^2+\deel{5\hbar}{48}(s_1-h)^2
\nonu
& & + \deel{\hbar}{6}s_2(s_1-h)-\deel{3\hbar}{16}s_2^2-
\deel{3\hbar^2}{8}s_2-\deel{7\hbar^2}{24}(s_1-h),
\nonu
C & = & (\deel{1}{2}ef+\deel{1}{2}fe+\deel{1}{4}h^2)+
(\deel{1}{4}s_2^2+\deel{\hbar}{2}s_2)+
(\deel{1}{8}s_1^2-\hbar s_1).
\ea
This completes our list of finite quantum $W$ algebras from
$sl_4$.

\noindent {\bf Acknowledgement}\\[7mm]
\noindent
We would like to thank Sander Bais and Jacob Goeree for many discussions
on the subjects presented in this paper and for reading the manuscript.
This work was financially supported
by the Stichting voor Fundamenteel Onderzoek der Materie (FOM).

\end{document}